\documentclass[lettersize,journal]{IEEEtran}
\usepackage{amsmath,amsfonts}
\usepackage{algorithmic}
\usepackage{comment}
\usepackage{algorithm}
\usepackage{array}
\usepackage[caption=false,font=scriptsize,labelfont=sf,textfont=sf]{subfig}
\usepackage{textcomp}
\usepackage{stfloats}
\usepackage{url}
\usepackage{verbatim}
\usepackage{graphicx}
\usepackage{cite}
\usepackage{bm} 
\usepackage{float}
\usepackage{amsmath} 
\usepackage{mathrsfs}
\usepackage{mathalpha}
\usepackage{booktabs}  
\usepackage{multirow}  
\usepackage{array}     
\usepackage{pifont} 
\usepackage[numbers,sort&compress]{natbib} 

\usepackage{indentfirst} 
\usepackage{titlesec}
\titlespacing*{\section}{0pt}{8pt}{6pt}
\titlespacing*{\subsection}{0pt}{6pt}{3.5pt}  

\DeclareMathOperator{\sigmoid}{sigmoid}
\newcommand{\sneq}{\mathrel{\mkern-5mu}=\mkern-5mu} 

\begin{document}
	
	\title{Synesthesia of Machines (SoM)-Empowered Wireless Image Transmission over Complex Dynamic Channel}
	\author{Haozhen Li,~\IEEEmembership{Graduate Student Member,~IEEE,} Ruide Zhang,~\IEEEmembership{Graduate Student Member,~IEEE,} \\Rongqing Zhang,~\IEEEmembership{Senior Member,~IEEE and} Xiang Cheng,~\IEEEmembership{Fellow,~IEEE}
		\thanks{This work was supported in part by the xxxx, xxxx, and xxxx.}  
		\thanks{Haozhen Li, Ruide Zhang and Xiang Cheng are with the State Key Laboratory of Photonics and Communications, School of Electronics, Peking University, Beijing 100871, China (email: pkuimlhz@pku.edu.cn; rdzhang25@stu.pku.edu.cn; xiangcheng@pku.edu.cn).}  
		\thanks{Rongqing Zhang is with Intelligent Transportation Thrust, The Hong Kong University of Science and Technology (Guangzhou), Guangzhou, China (e-mail: rongqingz@tongji.edu.cn).}  
	}
	
	
	\maketitle
	\begin{abstract}
		Wireless image transmission underpins diverse networked intelligent services and becomes an increasingly critical issue.
		Existing works have shown that deep learning-based joint source-channel coding (JSCC) is an effective framework to balance image transmission fidelity and data overhead.
		However, these studies oversimplify the communication system as a mere pipeline with noise, failing to account for the complex dynamics of wireless channels and concrete physical-layer transmission process.
		To address these limitations, we propose a Synesthesia of Machines (SoM)-empowered Dynamic Channel Adaptive Transmission (DCAT) scheme, designed for practical implementation in real communication scenarios.
		Building upon the Swin Transformer backbone, our DCAT scheme demonstrates robust adaptability to time-selective fading and channel aging effects by effectively utilizing the physical-layer transmission characteristics of wireless channels.
		Comprehensive experimental results confirm that DCAT consistently achieves superior performance compared with JSCC baseline approaches across all conditions.
		Furthermore, our neural network architecture demonstrates high scalability due to its interpretable design, offering substantial potential for cost-efficient deployment in practical applications.
	\end{abstract}
	
	\begin{IEEEkeywords}
		SoM, wireless image transmission, JSCC, time-selective fading, channel aging.
	\end{IEEEkeywords}
	
	\section{Introduction}\label{sec1}
	\IEEEPARstart{D}{riven} by emerging networked applications such as industrial Internet of Things (IoT), extended reality (XR), vehicular networks and autonomous driving (AD), beyond fifth generation (B5G), and sixth generation (6G) networks are dedicated to enabling “Intelligent Connectivity of Everything” with the powerful support of artificial intelligence (AI).
	These applications demand highly reliable, low-latency interactions capable of handling massive data volumes, with images playing a particularly critical role.  
	Images are ubiquitous in modern intelligent systems and contain rich semantic information, empowering a wide range of tasks including scene recognition \cite{Scene_recognition}, visual question answering \cite{VQA}, and object detection \cite{Object_detection}.
	Such significance necessitates capacity enhancement for image transmission in future wireless communication systems.
	
	Current image transmission systems are fundamentally based on the classical architecture of Shannon information theory \cite{Shannon1948}. Images are compressed into discrete symbols using coding protocols such as BPG, WebP, and JPEG, followed by channel coding to combat transmission errors, with LDPC \cite{LDPC} and Polar codes \cite{Polar} as typical examples widely adopted in 5G networks.
	In recent years, the advancement of deep learning has driven the progressive development of joint source-channel coding \textbf{(JSCC)} \cite{JSCC_2010} technology.
	With the help of neural networks, JSCC intends to realize efficient compression and transmission for specific data types like images, substantially reducing redundancy.
	It has been proven to achieve higher transmission efficiency and information fidelity under adverse channel conditions compared with separated architectures.
	DeepJSCC \cite{DeepJSCC}  employed the convolutional neural networks (CNN) for image compression and transmission, demonstrating the effectiveness of end-to-end JSCC implementation via deep neural networks. \cite{ResNet_ReID} and \cite{NTSCC} respectively investigated ResNet and Transformer as JSCC codecs.  
	With advancements in image processing networks, subsequent research \cite{ViT_JSCC_MIMO,WiTT} explored migrating Vision Transformer (ViT) \cite{ViT} and Swin Transformer \cite{SwinT} backbone network to further enhance image transmission performance.
	Generative AI techniques, including Generative Adversarial Network (GAN) and diffusion models, are also incorporated into recent JSCC frameworks \cite{GAN_JSCC,Diffusion_1,Diffusion_2}.
	Aligning with the “semantic” concept in Weaver's three-layer communication theory \cite{Weaver1953}, these methodologies are also commonly termed semantic communication \cite{SC_survey}.
	
	While the aforementioned works indeed accomplish efficient image transmission and successful support of downstream tasks with the help of AI tools, the majority of study efforts remain predominantly focused on image compression, neglecting the crucial importance of practical transmission system considerations.
	Such studies typically model the communication system as a mere pipeline with noise, which imposes great limitations in practical scenarios.
	Some emerging research begins to address these limitations, exploring the integration of JSCC with real-world wireless communication.
	For instance, \cite{WiTT}, \cite{ADJSCC}, and \cite{SNR} incorporated signal-to-noise ratio (SNR)-adaptive modules into the JSCC framework, to ensure that the model can work well across varying conditions.
	Some studies such as SwinJSCC \cite{SwinJSCC} and LIT \cite{LiT} further address dynamic transmission rates caused by system bandwidth variations by proposing targeted rate adaptation strategies.
	SCAN \cite{SCAN} specifically focused on the time-varying channel state information (CSI) in communication systems, enabling flexible CSI feedback while performing JSCC.
	
	Although these works preliminarily consider the dynamics of certain factors in wireless channels, the dynamic characteristics of the channel in practical systems are far more complex.
	Image transmission constitutes a non-instantaneous process with finite duration, mandating deep integration of the selective characteristics during image processing.  
	Besides, the concrete physical-layer transmission process exhibits more comprehensive dynamics, as there exists the need to obtain CSI through channel estimation \cite{ChannelEstimation_Pilot} and prediction \cite{WiFo,CSI_LLM}, which introduces nonideal issues like channel aging \cite{Aging}.
	Current studies exhibit deficiencies in this regard, lacking in-depth modeling and characterization of complex dynamic wireless channels. 
	Most simulation experiments are conducted under oversimplified channel models such as AWGN or Rayleigh fading and statistical parameter conditions, resulting in insufficient practicality for real-world systems.

	Against this backdrop, it is highly advisable to design a wireless image transmission system that transcends mere compressed representation, requiring tight coupling between image processing and practical physical-layer transmission characteristics.
	Inspired by human synesthesia, in which the stimulation of one sense organ will automatically evoke another sense organ to jointly perform cognitive tasks, the Synesthesia of Machines (SoM) is proposed in \cite{SoM}. It aims to leverage artificial neural networks to extract high-density, task-aware, and robust features, thereby achieving intelligent integration of communication and multi-modal sensing \cite{SoM_TNSE}. Guided by the SoM paradigm, our goal is to deeply couple image transmission tasks with complex dynamic channel conditions to develop a more practical solution. To address existing challenges, our target scheme is expected to be compatible with actual communication systems and robust enough against various non-ideal factors.
	
	Therefore, we present a SoM-empowered wireless image transmission scheme \textbf{Dynamic Channel Adaptive Transmission}, referred to as \textbf{DCAT}. 
	We perform extensive experimental evaluations to confirm that it outperforms existing JSCC-based wireless image transmission approaches across different channel conditions.
	In practical situations, our proposed DCAT achieves at least 10.5\% PSNR improvement and 12.6\% LPIPS reduction while maintaining identical communication overhead, and demonstrates superior scalability across diverse scenarios.
	Accordingly, DCAT is able to achieve more efficient and high-fidelity image transmission and support emerging applications such as IoT and intelligent transportation systems. The key contributions of our work are in the following aspects:
	
	\begin{itemize}
	\item To tackle the reliability challenges posed by complex non-ideal channel conditions in image transmission systems, we propose a tightly-coupled architecture bridging visual image processing with wireless transmission under the guidance of the SoM paradigm, demonstrating outstanding practicality in real-world communication scenarios.
	We employ a three-stage training strategy to holistically learn multi-domain knowledge, enabling the adaptation of image feature coding to dynamic wireless channels.
	
	\item To empower DCAT with flexible capabilities for diverse complex scenarios, we effectively utilize physical-layer transmission information of actual communication systems as side information to enhance the image codecs.
	Unlike existing JSCC approaches that conduct experiments solely based on simplified AWGN or Rayleigh channels, we consider realistic channel conditions and correspondingly design two efficient adaptive modules, DC-attn and DC-permu for targeted optimization.
	Consequently, our DCAT exhibits great adaptability to the time-selective fading channel, especially exceptional robustness against channel aging effects.
	
	\item To enhance the efficiency and deployment potential, we design an interpretable neural network by leveraging rule-based domain knowledge in wireless communications, controlling reasonable computing overhead.
	Moreover, with efficient parameter fine-tuning, DCAT achieves superior scalability, enabling cost-effective migration and deployment in diverse networked applications.
	\end{itemize}
	
	The remainder of this paper is structured as follows. 
	Sec. \ref{sec2}. formally presents the problem formulation of image transmission over complex dynamic wireless channels and gives the system model. 
	Then, Sec. \ref{sec3}. elaborates on
	the proposed DCAT methodology. 
	Sec. \ref{sec4}. shows the details of the experimental setup.
	Further, Sec. \ref{sec5}. provides comprehensive simulation results to validate the efficacy of our proposed DCAT. 
	Finally, Sec. \ref{sec6}. highlights our research conclusions.

	\section{Scenario and System Model}\label{sec2}
	This study focuses on image transmission in practical communication systems over dynamic wireless channels.
	The dynamic channel under our consideration refers specifically to the time-selective fading channel with aging effects.
	\subsection{Time-selective fading channel with aging} \label{sec2-1}
	In contrast to the oversimplified channel models considered in existing studies, real-world communication systems exhibit complex and highly dynamic channel characteristics.
	The inherent randomness of electromagnetic wave propagation, combined with the dynamic nature of the environment, particularly the mobility of the transceiver, results in pronounced time-selectivity.
	For a single-carrier, single-input single-output scenario, the channel model can be formulated as:
	\begin{equation}
		\label{basic_channel}
		Y(t) = H(t) \cdot X(t)+N(t)
	\end{equation}
	where $X(t) \in \mathbb{C}$ and $Y(t)\in \mathbb{C}$ denote the transmitted and received symbol at time instant $t$ respectively. $\mathbb{C}$ denotes the set of complex numbers.
	The transmitted signal $\bm{X}$  is subject to a power constraint $P_s$:
	\begin{equation}
		\label{power_constraint}
		E_s=\mathbb{E} [\|\bm{X}\|_2^2] \leq P_s
	\end{equation}
	where $ \| \cdot \|_2 $ denotes the Euclidean Norm.
	$N(t) \in \mathbb{C}$ is additive noise. Each element is sampled from independent and identically distributed complex Gaussian noise as $\bm{N} \sim \mathcal{CN}(0,\sigma_n^2)$.
	
	In Eq. (\ref{basic_channel}), it is crucial to emphasize that \( H(t) \) denotes the time-varying CSI in real-world environments.
	Without loss of generality, we can perform power normalization to maintain a consistent SNR between the transceiver, i.e., $\mathbb{E} [\|\bm{HX}\|_2^2] = \mathbb{E} [\|\bm{X}\|_2^2]  = E_s$.
	Correspondingly, the SNR can be defined as:
	\begin{equation}
		\label{SNR_define}
		\text{SNR}=10\log_{10}\frac{E_s}{E_n}=10 \log_{10}\frac{E_s}{\sigma_n^2} \text{ (dB)}
	\end{equation}

	\begin{figure}[!t] 
		\centering
		\subfloat[]{
			\includegraphics[width=0.85\linewidth]{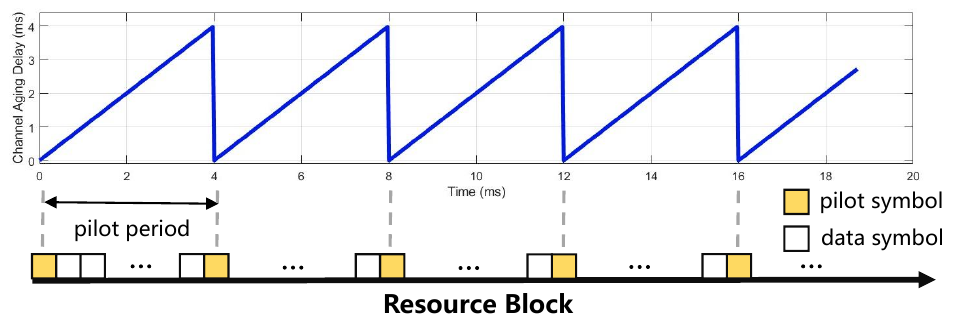} 
			\label{pic_aging:subfig_a}
		}
		\vspace{0.01cm} 
		\subfloat[]{
			\includegraphics[width=0.46\linewidth]{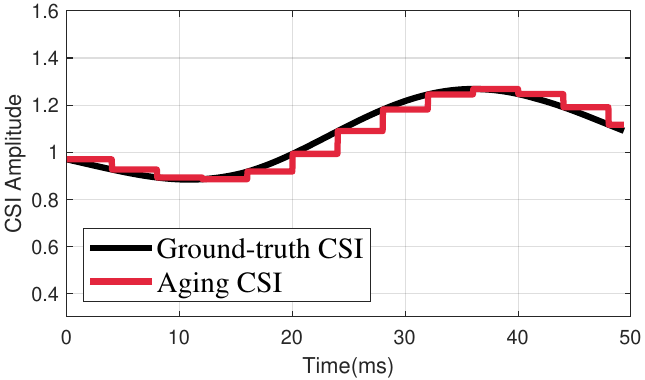} 
			\label{pic_aging:subfig_b}
		}
		\hfill 
		\subfloat[]{
			\includegraphics[width=0.46\linewidth]{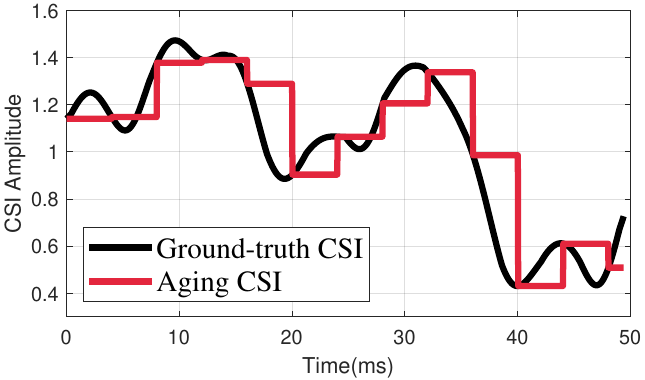} 
			\label{pic_aging:subfig_c}
		}
		\caption{Illustration of channel aging. 
			(a) Aging delay induced by periodic pilot symbols;
			(b) CSI distortion caused by channel aging at $v=2$ m/s;
			(c) CSI distortion caused by channel aging at $v=15$ m/s.}
		\label{pic_aging}
	\end{figure}

	CSI fundamentally represents the channel frequency response, which forms a Fourier transform pair with the channel impulse response characterized by the following expression:
	\begin{equation}
		\label{cir}
		h(t,\tau) = \sum_{l=1}^{L_p} a_l(t) e^{-j\phi_l(t)} \delta(\tau - \tau_l(t))  
	\end{equation}
	where $L_p$ is the number of multipath propagation components.
	$a_l(t)$,  $\phi_l(t)$, and $\tau_l(t)$ denote the amplitude, phase, and delay, respectively.
	The phase component is particularly significant because it is highly sensitive to channel dynamics, as illustrated below. Such sensitivity directly contributes to the time-selective fading observed in wireless channels.
	
	\begin{equation}
		\label{phase_define}
		\phi_l(t)=2\pi f_c \frac{d_l}{c}+2\pi f_c\frac{v t \cos(\theta_l)}{c}
	\end{equation}
	$c$ is the velocity of light, $f_c$ is the carrier frequency, $d_l$ and $\theta_l$ represent the propagation distance and direction of the corresponding path, respectively.
	Crucially, $v$ denotes the velocity between the transceiver, introducing Doppler rotation as the primary cause of time-selective fading.
	Moreover, the receiver must acquire the real-time CSI to perform channel equalization, where we use zero-forcing (ZF) equalization as:
	\begin{equation}
		\label{ZF}
		\hat{X}(t) = \frac{Y(t)}{\hat{H}(t)}
	\end{equation}
	where $\hat{X}(t)$ denotes the equalized symbol, $\hat{H}(t)$ denotes the CSI estimates acquired via channel estimation.
	It should be noted that we consider narrowband channel where inter-symbol interference (ISI) is absent. Under this circumstance, the ZF equalization can be formulated as Eq. (\ref{ZF}), which processes each symbol independently at each time instant $t$.
	Meanwhile, as discussed in Sec. \ref{sec1}., in actual systems what generally exists are \textbf{\textit{Aging Scenarios}}. Channel Aging phenomenon arises because channel estimation relies on periodically transmitted pilot signals, which cannot provide perfectly accurate and real-time CSI value. We specifically consider the following two situations:
	\begin{equation}  
		\label{two_cases}
		\hat{H}(t) =\begin{cases}
			H(t),   &  \textit{Aging Scenario with CP} \\
			H(t-\tau_{ag}(t)),   &  \textit{Aging Scenario}\\
		\end{cases}               
	\end{equation}
	
	The former represents an idealized situation, where we assume that the CSI at any time instant can be acquired without error through physical-layer \textbf{channel prediction (CP)} techniques.
	While under high-mobility situations causing severe CP failure, the latter situation aligns more closely with practical reality. Only at the moment corresponding to the pilot signal can the accurate CSI be acquired. 
	$\bm{\tau_{ag}}$ denotes the time interval since the previous pilot symbol.
	As shown in Fig. \ref{pic_aging}, $\tau_{ag}(t)$ exhibits a characteristic sawtooth pattern over time $t$, which can be formally expressed by the following equation:
	\begin{equation}   
		\label{aging_time}
		\tau_{ag}(t)= t- \lfloor \frac{t}{T_p} \rfloor T_p
	\end{equation}
	where $ T_p $ denotes the pilot interval period.
	For a fixed pilot period, CSI distortion grows with the intensity of time-selectivity. 
	
	\subsection{Wireless image transmission over dynamic channel}\label{sec2-2}
	We consider a point-to-point image transmission system with physical-layer transmission information feedback as shown in Fig. \ref{pic_system_model}. 
	During the operation of the image transmission system, the transmitter (Tx) and receiver (Rx) can get the relevant information of the concrete physical-layer transmission which comprises the following key components:
	\begin{equation}   
		\label{Transmisson Infomation}
		\bm{\mathcal{P}}= \Big\{ \text{SNR}, \bm{\hat{H}}, \overline{f_D},  \bm{\tau_{ag}} \Big\}  
	\end{equation}
	where we name $\bm{\mathcal{P}}$ as “Physical-Layer Transmission Information” with detailed explanations below:
	
	\begin{figure}[!t]
		\centering
		\includegraphics[width=0.92\linewidth]{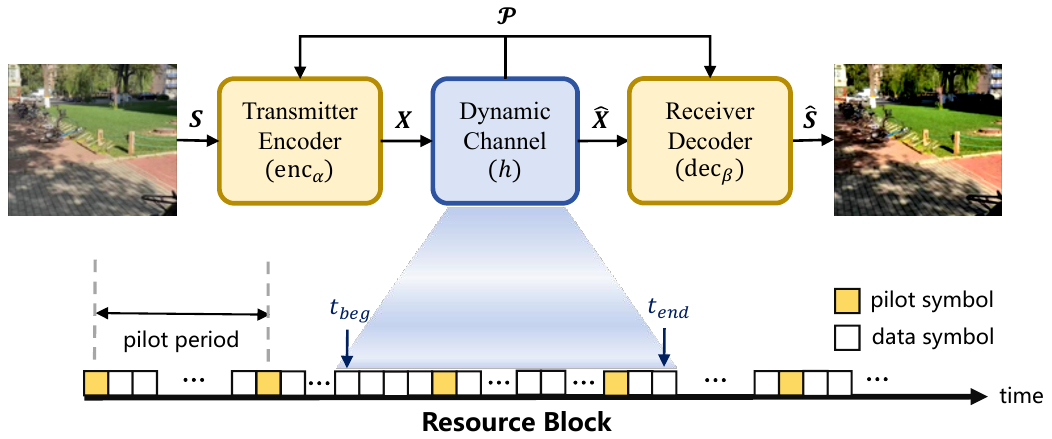}
		\caption{The model of the point-to-point image transmission system with physical-layer transmission information $\bm{\mathcal{P}}$ feedback. }
		\label{pic_system_model}
	\end{figure}

	\begin{itemize}
		\item $\text{SNR}$: Since the duration of image transmission is significantly shorter than the coherence time of the SNR variation, it can be approximated as a constant value during a single image transmission session.
		\item $\bm{\hat{H}}$: This term denotes the CSI estimates available for the transceiver as formulated in Eq. (\ref{two_cases}).
		\item $\overline{f_D}$: It represents the normalized Doppler value  and reflects the intensity of time-selective fading as $\overline{f_D}= \lambda \frac{f_c v}{c}$,
		where $\lambda$ is a constant scaling factor.
		It can also be treated as a scalar quantity since it remains nearly constant during the transmission time window of a single image.
		\item $\bm{\tau_{ag}}$: It denotes the channel aging delay which can be determined from the physical layer.
	\end{itemize}
	Based on the above analysis, we formulate the wireless image transmission problem over the dynamic channel as follows.
	
	Our system focuses on transmitting a single $m$-channel image $\bm{S} \in \mathbb{R}^{H_0 \times W_0 \times m}$ as the basic unit, where $\mathbb{R}$ denotes the set of real numbers, $H_0$ and $W_0$ denote the height and width of the image. The transmission process can be divided into four sequential steps:
	
	\subsubsection{Physical-Layer Infomation Feedback}\label{sec2-2-1}
	At the moment $t_{beg}$ image transmission begins, the Tx-side encoder needs to obtain the latest feedback information $\bm{\mathcal{P}}$.
	The information available for the Tx can be presented as $\bm{\mathcal{P}}_{tx} \sneq \Big\{ \text{SNR},  \bm{\hat{H}}(:t_{beg}), \overline{f_D},\bm{\tau_{ag}} \Big\}  $,
	which serves as side information for the encoder. 
	$\bm{\hat{H}}(:t_{beg}) $ denotes the available CSI vector before $t_{beg}$, which guarantees the causality of the system.

	\subsubsection{Encoding at the Transmitter}\label{sec2-2-2}
	Based on the original image $\bm{S}$ and the side information ${\bm{\mathcal{P}}_{tx}}$ obtained in the previous step, the encoded
	feature can be represented by:
	\begin{equation}   
		\label{encoder_formula}
		\bm{z}=\mathrm{enc}_\alpha(\bm{S},\bm{\mathcal{P}}_{tx})   
	\end{equation}
	where $\bm{z} \in \mathbb{R}^{L \times C}$ denotes a token sequence with length $L$ and channel dimension $C$. 
	To avoid ambiguity with the wireless channel or physical channel,
	we refer to the channel here as “feature-channel”.
	$\mathrm{enc}_\alpha$ represents the encoder network with the parameter set $\alpha$. 
	The image Compression Ratio (CR) can be defined as:
	\begin{equation}   
		\label{CR_define}
		\text{CR}=\frac{L \times C}{H_0 \times W_0 \times m} 
	\end{equation}

	\subsubsection{Symbol Transmission over the Physical Channel} \label{sec2-2-3}
	Practical wireless communication systems transmit constellation symbols as the basic unit. We need to modulate the token sequence $\bm{z}$:
	\begin{equation}   
		\label{MOD}
		\bm{X}=\text{MOD}(\bm{z})
	\end{equation}
	Specifically,  the first $\frac{C}{2}$ feature-channels are treated as the real component, while the remaining $\frac{C}{2}$  form the imaginary component.
	This complex representation yields a total of $n=\frac{L\times C}{2}$ symbols, which are then temporally ordered through a flattening operation that prioritizes the feature-channel dimension over the token dimension.
	During the time window from $t_{beg}$ to $t_{end}$, these symbols are sequentially transmitted through the channel described in Sec. \ref{sec2-1}.:
	\begin{equation}   
		\label{Trans_in_wireless}
		\bm{\hat{X}}=\bm{\hat{H}}^{-1}(\bm{H}\bm{X}+\bm{N})
	\end{equation}
	At the Rx, the received symbols undergo a demodulation process that is the inverse of the modulation operation, i.e.,
	\begin{equation}   
		\label{DEMOD}
		\bm{\widetilde{z}}=\text{DEMOD}(\bm{\hat{X}})
	\end{equation}
	For end-to-end training of the model, all the operations should be differentiable that allow back-propagation.
	Therefore, we utilize the straight-through estimator (STE) \cite{STE} to model the impact of  composition of transformations in Eqs. (\ref{MOD})(\ref{Trans_in_wireless})(\ref{DEMOD}) as equivalent perturbation. 
	The overall process of symbol modulation, transmission over the physical channel, and demodulation is represented by the function $h$:
	\begin{equation}   
		\label{STE}
		\bm{\hat{z}}=h(\bm{z})=\bm{z}+(\bm{\widetilde{z}}-\bm{z})_{detach}
	\end{equation}
	where $detach$ denotes the removal of gradients in the computational graph of the network.
	When computing the gradient of the loss funcition $Loss$ during backpropagation, we have:
	\begin{equation}   
		\label{STE_loss_func}
		\frac{\partial Loss}{\partial \bm{z}}=\frac{\partial Loss}{\partial \bm{\hat{z}}}
	\end{equation}

	\subsubsection{Decoding at the Receiver}\label{sec2-2-4}
	Upon completion of symbol transmission, the Rx initiates image reconstruction at $t_{end}$.
	Similar to the encoder, the decoder also needs to utilize $\bm{\mathcal{P}}$ as side information for reconstructing the image.
	Owing to the temporal disparity, the Rx is able to acquire more recent physical-layer information, which can be represented by 
	$\bm{\mathcal{P}}_{rx} \sneq \Big\{ \text{SNR},  \bm{\hat{H}}(:t_{end}), \overline{f_D},\bm{\tau_{ag}} \Big\}$.
	The decoder can be expressed as:
	\begin{equation}   
		\label{decoder_formula}
		\bm{\hat{S}}=\mathrm{dec}_\beta(\bm{\hat{z}},\bm{\mathcal{P}}_{rx})   
	\end{equation}
	where $\bm{\hat{S}} \in \mathbb{R}^{H_0 \times W_0 \times m}$ denotes the reconstructed image and $\mathrm{dec}_\beta$ is the decoder network with the parameter set $\beta$.
	
	Integrating all the aforementioned steps, our system aims to optimize the encoder and decoder to minimize the distortion between the original image and the reconstructed image given the constraint of CR across all channel conditions, which can be formulated by:
	\begin{equation}   
		\label{overall_formula}
		\begin{split}
			(\alpha^{\ast},\beta^{\ast})=&\mathop{\arg\min}_{\alpha,\beta} d(\bm{S},\bm{\hat{S}}) \\
			s.t. \quad \bm{\hat{S}}=\mathrm{dec}_\beta&(h(\mathrm{enc}_\alpha(\bm{S},\bm{\mathcal{P}}_{tx})),\bm{\mathcal{P}}_{rx})
		\end{split}		
	\end{equation}
	where $d(\cdot)$ denotes the distortion metric function.

	\section{SoM-Empowered Dynamic Channel Adaptive Transmission Scheme Design}\label{sec3}

\begin{figure*}[!t]
		\centering
		\includegraphics[width=0.78\linewidth]{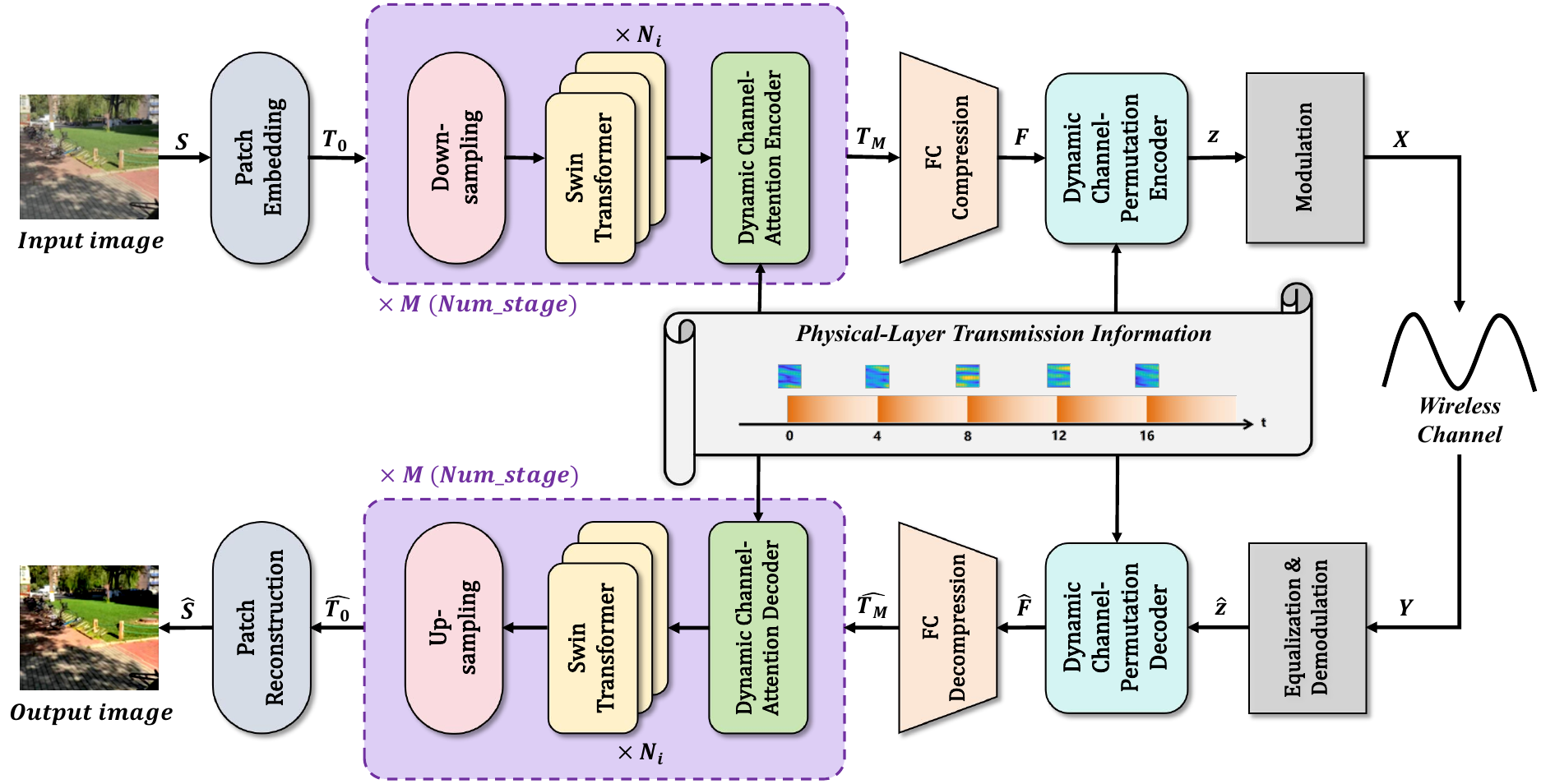}
		\caption{Overall architecture of DCAT, our proposed wireless image transmission scheme over the complex dynamic channel.}
		\label{pic-method-overall}
	\end{figure*}

	In order to achieve an efficient and robust wireless image transmission over complex dynamic scenarios, we propose a Dynamic Channel Adaptive Transmission \textbf{(DCAT)} framework.
	The overall architecture of the scheme is illustrated in Fig. \ref{pic-method-overall}.
	We adopt the state-of-the-art (SOTA) architecture in image processing, Swin Transformer (SwinT) \cite{SwinT}, as our backbone. On this foundation, we propose \textbf{Dynamic Channel-Attention} (DC-attn) and \textbf{Dynamic Channel-Permutation} (DC-permu) modules to flexibly adapt to transmission impairments from real-world non-ideal channel conditions.
	Through the combined effect of the image processing backbone and the dynamic channel adaptation components, DCAT enables wireless channel-adaptive image compression and reconstruction process.
	Moreover, we utilize a transfer learning strategy to optimally balance knowledge between image feature processing and wireless channel adaptation, which can boost training efficiency and build a reliable neural network model.
	
	\subsection{Overall Workflow based on Swin Transformer BackBone}\label{sec3-1}
	SwinT enhances the Transformer architecture by introducing a shifted window mechanism.
	Distinct from conventional natural language processing (NLP)-oriented Transformers \cite{All_you_need}, it specifically addresses vision requirements through  Windows Multi-Head Self-Attention (W-MSA) and Shifted Windows Multi-Head Self-Attention (SW-MSA), which facilitates more efficient feature extraction.   
	This CNN-liked hierarchical structure delivers superior performance across various vision tasks.
	\begin{figure}[!t]
		\centering
		\includegraphics[width=0.87\linewidth]{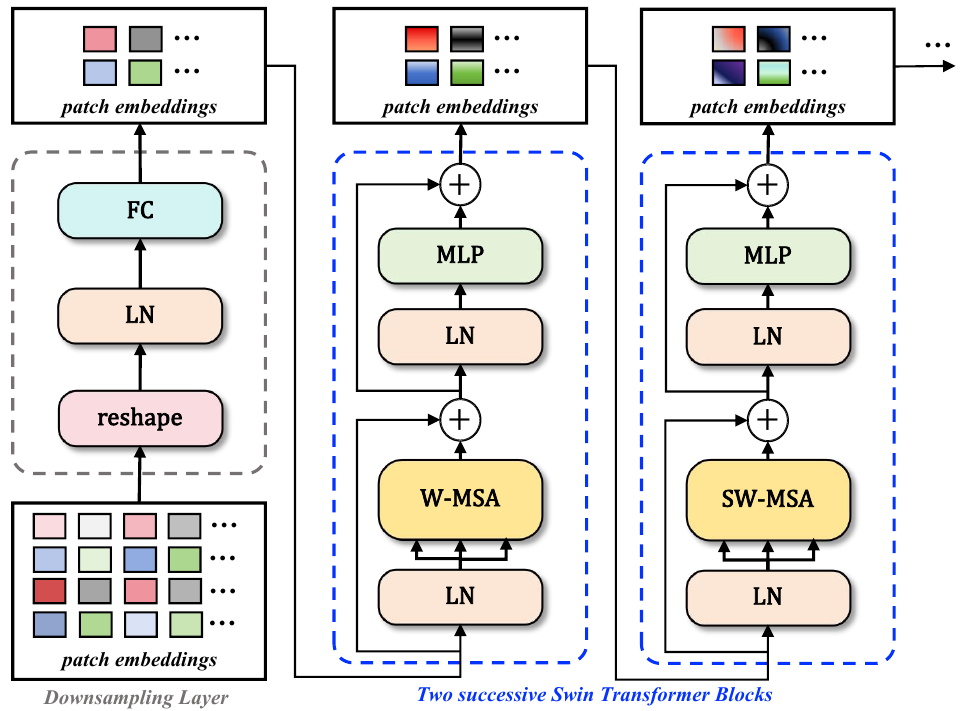}
		\caption{The processing flow within a single stage of the encoder. SwinT blocks typically operate in pairs, alternately executing W-MSA and SW-MSA.}
		\label{pic-swint}
	\end{figure}
	
	When combined with simple linear sampling layers, it can well empower the image transmission task. 
	As shown in Fig. \ref{pic-swint}, the input image $\bm{S}$ is transformed into patches via a convolutional layer, which are then represented as a token sequence $\bm{T_0}$.
	This is followed by $M$ stages, where the $i$-th stage firstly applies fully-connected (FC) downsampling to reduce the number of tokens, then processes them through $N_i$ stacked Transformer blocks for feature transformation, gradually compressing the image.
	The output of the $i$-th stage is $\bm{T_i} \in \mathbb{R}^{L_i \times C_i} $ ($L_i=\frac{H_0 W_0}{2^{2i}}$), where each stage reduces the token count by $4 \times$. To further reduce data volume burden, an FC compression module is applied to reduce the number of feature-channels to $C$. The final CR can be expressed as $\frac{C}{m \times 4^M}$ according to Eq. (\ref{CR_define}).
	
	The Rx employs the symmetric structure, utilizing FC decompression to increase feature-channel dimensionality of the equalized feature representation $\bm{\hat{z}}$, followed by $M$ stages for progressive feature decoding. Finally, the patch tokens are reshaped to output the reconstructed image $\bm{\hat{S}}$.
	
	However, the aforementioned processing focuses solely on data compression and decompression.    
	In order to enable adaptive capability for real transmission  conditions, we need to effectively utilize the corresponding information.
	The following two sections introduce the DC-attn and DC-permu components, which serve as dimension-preserving modules that seamlessly integrate into the existing workflow, collaboratively improving the performance of our image transmission system.
	
	\subsection{Dynamic Channel-Attention (DC-attn) Modules }\label{sec3-2}

	In wireless image transmission, the physical-layer transmission information $\bm{\mathcal{P}}$ can serve as side information to enhance image processing.
	We employ attention mechanisms at each processing stage of both the Tx and Rx to adaptively refine image features. Our designed DC-attn operates in conjunction with Transformer blocks, progressively realizing adaptive matching of image features to transmission conditions.
	
\begin{figure}[!t]
		\centering
		\includegraphics[width=0.9\linewidth]{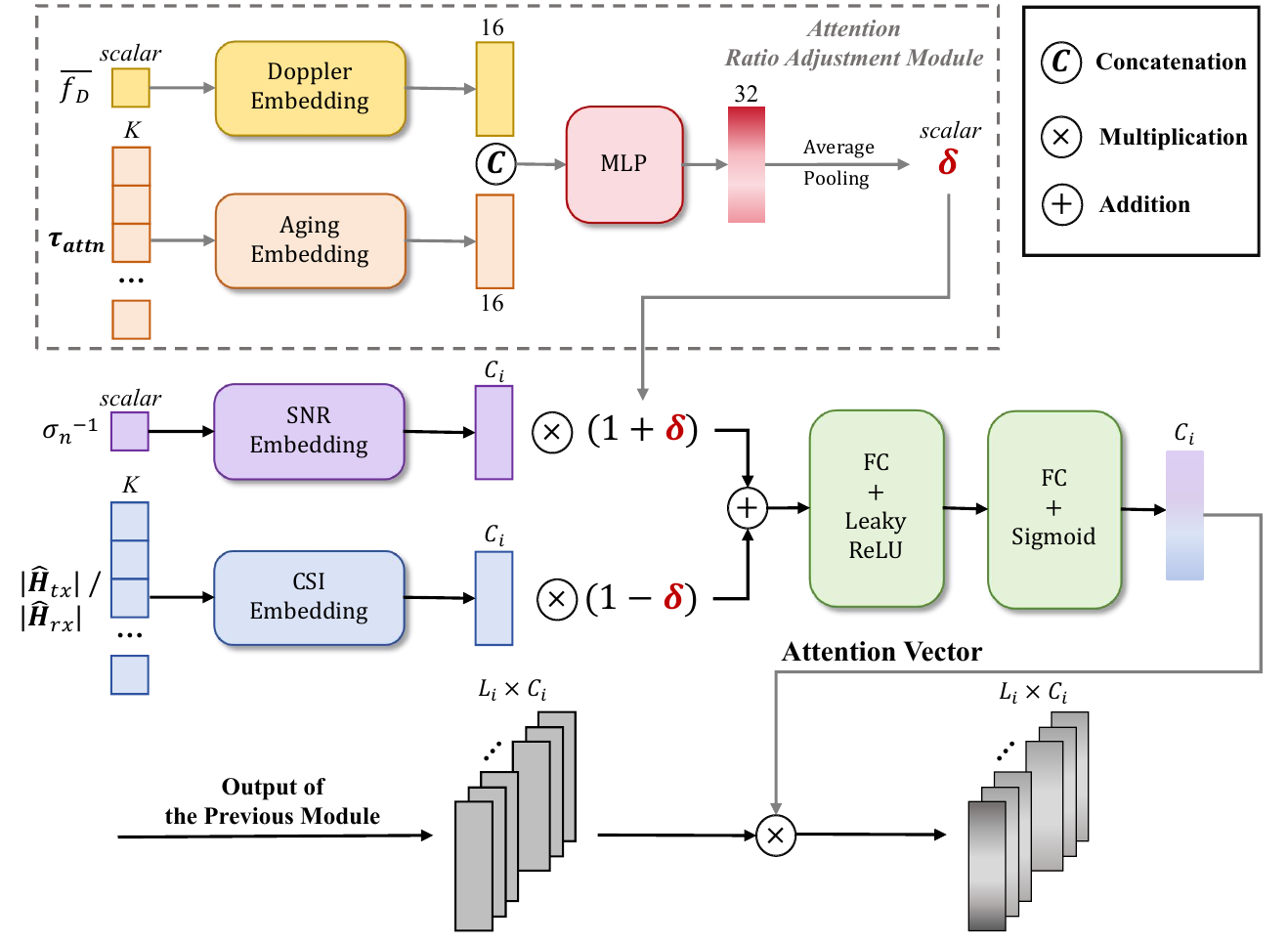}
		\caption{Diagram of the Dynamic Channel-Attention modules. Each embedding submodule is an MLP employing Leaky ReLU for hidden layer activation, to transform the side information into distinct feature representations.}
		\label{pic-DC-attn}
	\end{figure}

	As depicted in Fig. \ref{pic-DC-attn}, at the $i$-th stage ($1 \le i \le M$), we utilize side information to generate an \textbf{Attention Vector} of dimension $C_i$, which serves as a scaling factor applied to all $L_i$ tokens, globally modulating the weights across different feature-channels.
	Our methodology thoroughly incorporates the physical significance of these pieces of information. SNR and $\bm{\hat{H}}_{tx}$/$\bm{\hat{H}}_{rx}$ characterize the overall channel conditions experienced during image transmission. The image transmission process spans $K$ time units from $t_{beg}$ to $t_{end}$. 
	According to Sec. \ref{sec2-2-3}., these $K$ units correspond to the transmission of $C$ feature dimensions through the wireless channel via real and imaginary components, thereby yielding $K=C/2$.
	The CSI embedding submodule converts the complete amplitude profile over this period into feature representations, which are then combined with SNR-derived features and further transformed into the Attention Vector normalized to $(0,1)$.
	It is crucial to emphasize that under \textit{Aging Scenario}, due to the dynamic nature of wireless channels, we introduce an additional \textbf{Attention Ratio Adjustment (ARA) Module} to deal with the uncertainty in the input $\bm{\hat{H}}_{tx}$/$\bm{\hat{H}}_{rx}$:
	\begin{equation}   
		\label{encoder_decoder_csi_att}
		\bm{\hat{H}}_{tx}(t) = \bm{\hat{H}}(t_{beg}) \text{, } \bm{\hat{H}}_{rx}(t) = \bm{\hat{H}}(t)        
	\end{equation}

	The distinction between expressions of two sides arises from considerations of system causality. In other words, when performing computations at any given time instant, we can only utilize available information before that moment.
	To this end, $\bm{\tau_{attn}}$ represents the time difference from the actual transmission moment at each time step, derived as:
	\begin{equation}   
		\label{encoder_decoder_tau_att}
		\tau_{attn}[k] =\begin{cases}
			\tau_{ag}(t_{beg})+(t_{beg}-\mathcal{T}[k]),   &  \text{for the Tx} \\
			\tau_{ag}(\mathcal{T}[k]),   &  \text{for the Rx}\\
		\end{cases}               
	\end{equation}
	where the $\bm{\tau_{ag}}$ is determined by Eq. (\ref{aging_time}) and $k$ denotes each discrete moment. $\mathcal{T}[k]$ specifies the physical time corresponding to index $k$ by $\mathcal{T}[k]=t_{beg}+\frac{k-1}{K-1}(t_{end}-t_{beg})$ ($1 \le k \le K$).

	The ARA Module generates a confidence factor $\bm{\delta}$ through the embedding of $\bm{\tau_{attn}}$ and $\overline{f_D}$, which exhibits positive correlation with the uncertainty of $\bm{\hat{H}}_{tx}$/$\bm{\hat{H}}_{rx}$ and serves to modulate feature weighting proportions.
	We introduce an additional regularization term to optimize the parameters of the ARA Module, ensuring the adjustment effect aligns with our design objectives. For implementation, we adopt the Normalized Mean Square Error (NMSE) of CSI, a widely adopted metric in channel estimation and prediction field \cite{CSI_NMSE}:
	\begin{equation}   
		\label{CSI_NMSE}
		\text{NMSE}(\bm{\hat{H}}_{tx/rx},\bm{H}_{gt})=\frac{\| \bm{\hat{H}}_{tx/rx}-\bm{H}_{gt} \|_2^2}{\| \bm{H}_{gt} \|_2^2}   
	\end{equation}
	where $\bm{H}_{gt}$ denotes the ground truth of CSI. We utilize the NMSE of CSI within the corresponding interval to optimize the confidence factor $\bm{\delta}$ output, incorporating the following expression as regularization terms into the training loss function:
	\begin{equation}   
		\label{regularization_term_nmse}
		L_{attn\_tx/rx}=|\bm{\delta}_{tx/rx}-\text{NMSE}(\bm{\hat{H}}_{tx/rx},\bm{H}_{gt})|
	\end{equation}

	\subsection{Dynamic Channel-Permutation (DC-permu) Modules }\label{sec3-3}
	\begin{figure*}[!t]
		\centering
		\includegraphics[width=0.84\linewidth]{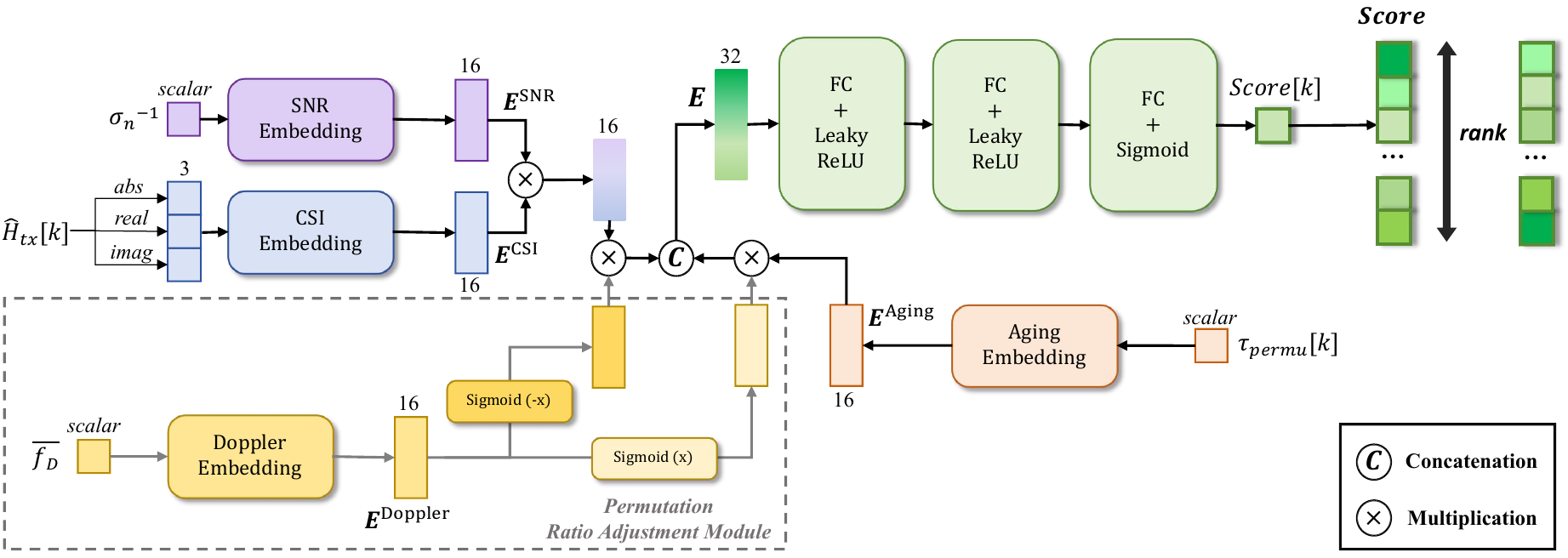}
		\caption{Diagram of the Dynamic Channel-Permutation modules. Each embedding submodule is an MLP employing Leaky ReLU for hidden layer activation. The side information at different time instant $k$ generates the scoring result with shared parameters, which facilitate feature-channel permutation.}
		\label{pic-DC-permu}
	\end{figure*}
	DC-attn can globally integrate complex dynamic nature through attention-weighted feature-channels, effectively coupling image features with side information from wireless channels.
	Meanwhile, we propose the Dynamic Channel-Permutation (DC-permu) module to process different feature-channels at a finer granularity, operating in complementary coordination with DC-attn and further enhancing the robustness of the transmission system against dynamic channels.
	
	DC-permu adopts a fully symmetric structure, executing mutually inverse transformations in the encoder and decoder to achieve feature-channel permutation and restoration. At the Tx, through stage-wise token merging and FC-layer-based feature compression, it forms a compressed representation $\bm{F} \in \mathbb{R}^{L \times C}$ ($L=\frac{H_0 W_0}{4^M}$) of the image $\bm{S}$. The $C$-dimensional feature-channels are completely parallel, but their contributions to the task might vary. That means
	certain feature-channels may contain more critical information while others carry relatively secondary information. This characteristic precisely aligns with the time-selective nature and channel aging of the dynamic channel. 
	Under the guidance of the aforementioned theory, we perform feature-channel permutation prior to symbol modulation at the Tx, while executing inverse permutation after equalization and demodulation at the Rx, as expressed by:
	\begin{equation}   
		\label{permutation}
		\begin{split}
			\bm{z}&=\Psi({\bm{F}})=[F_{\psi(1)},F_{\psi(2)},…,F_{\psi(C)}] \\
			\bm{\hat{F}}&=\Psi^{-1}(\bm{\hat{z}})=[\hat{z}_{\psi^{-1}(1)},\hat{z}_{\psi^{-1}(2)},…,\hat{z}_{\psi^{-1}(C)}]
		\end{split}
	\end{equation}
	where $\psi$ and $\psi^{-1}$ denote a pair of reversible permutations for the index set $[C]$.
	This approach facilitates more granular utilization of the feature, significantly improving the robustness against dynamic channel conditions.
	
	In practical implementation, DC-permu Encoder and Decoder employ the network architecture with shared parameters shown in Fig. \ref{pic-DC-permu}, for generating the permutation rule $\psi$ and $\psi^{-1}$.
	Unlike DC-attn which processes side information from all $K$ time units holistically, DC-permu generates a distinct $Score[k]$ for each discrete time instant $k$. These scores are sorted in ascending order, where smaller values indicate better instantaneous channel quality. The corresponding real and imaginary components of the $K \sneq C/2$ feature-channels are then permuted accordingly.
	
	As illustrated in Fig. \ref{pic-DC-permu}, we denote the 16-D feature tensors generated from side information $\bm{\mathcal{P}}$ as $\bm{E}^{\text{SNR}}$, $\bm{E}^{\text{CSI}}$, $\bm{E}^{\text{Doppler}}$, $\bm{E}^{\text{Aging}}$, respectively. The final scoring result can be gotten from the joint representation vector $\bm{E}$:
	\begin{equation}   
		\label{score}
		\begin{split}
		\bm{E}_{1}= \bm{E}^{\text{SNR}} \odot & \bm{E}^{\text{CSI}} \odot \sigmoid (-\bm{E}^{\text{Doppler}})\\
	    \bm{E}_{2}= \bm{E}^{\text{Aging}}  & \odot \sigmoid(\bm{E}^{\text{Doppler}})\\
		\bm{E}=& Concat(\bm{E}_{1},\bm{E}_{2})
		\end{split}
	\end{equation}
	where $\odot$ is the channel-wise feature multiplication operation.
	$\bm{E}_{1}$ represents the noise level estimation during transmission, while  $\bm{E}_{2}$ specifically accounts for the impact of channel aging on symbol equalization.
	$\bm{E}^{\text{Doppler}}$ reflects the magnitude of time-selectivity and is used to adjust the weight between $\bm{E}_{1}$ and $\bm{E}_{2}$ based on \textbf{Permutation Ratio Adjustment Module}. 
	
	For \textit{Aging Scenario with CP} we have: $\bm{\hat{H}}_{tx} \sneq \bm{\hat{H}}_{rx} \sneq \bm{H}$, $\bm{\tau_{permu}} \sneq 0$, thereby $\bm{E}_{2}$ becomes inactive. For \textit{Aging Scenario}, the expressions for $\bm{\hat{H}}_{tx}$ and $\bm{\tau_{permu}}$ are given by:
	\begin{equation}   
		\label{permu_def}
		\hat{H}_{tx}[k]= \hat{H}(\mathcal{T}[1]) \text{, }
		\tau_{permu}[k]= \tau_{ag}(\mathcal{T}[k])
	\end{equation}
	where two critical points warrant emphasis. First, the CSI estimates exclusively utilize $\bm{\hat{H}}_{tx}$ from the Tx-side to ensure the equal inputs for the DC-permu Encoder and Decoder. Only under this condition can identical scoring results be generated, enabling permutation and restoration without sharing any additional information. Second, the physical significance of $\bm{\tau_{permu}}$ differs from that of $\bm{\tau_{attn}}$. 
	It reflects the aging delay of each feature-channel during channel equalization.
	
	Given that the sorting operation is non-differentiable, we introduce a regularization term similar to DC-attn to provide additional knowledge for the model.
	With the goal that the scoring results can well reflect the symbol transmission impairment magnitude for corresponding indices, we define:
	\begin{equation}   
		\label{index_impairment}
			\bm{Imp}(\bm{z},\bm{\hat{z}})=\tanh(\| \bm{z}_{re}-\bm{\hat{z}}_{re} \|_2^2+\| \bm{z}_{im}-\bm{\hat{z}}_{im} \|_2^2)
	\end{equation}
	where $\bm{z}_{re}$ and $\bm{z}_{im}$ denote the first and last $K$ feature-channels of $\bm{z}$ (i.e., the real and imaginary components), with $\bm{\hat{z}}$ following the same convention. We employ the nonlinear $\tanh$ function to map these values to $(0,1)$ range.
	We incorporate the following regularization term into the loss function to provide supervisory constraints on the DC-permu scoring output:
	\begin{equation}   
		\label{regularization_term_tanh_mse}
			L_{permu}=|\bm{Score}-\bm{Imp}(\bm{z},\bm{\hat{z}})|
	\end{equation}
	\subsection{Training Strategy}\label{sec3-4}
	Integrating all above, we define the loss function to support neural network training as:
	\begin{equation}   
		\label{total_loss}
		Loss=L_{image}+\omega_1 L_{attn\_tx} + \omega_2 L_{attn\_rx} + \eta L_{permu}
	\end{equation}
	where $L_{image}$ quantifies image perceptual distortion and is closely related to the performance evaluation of image transmission, which will be discussed in Sec.\ref{sec4-2}. The hyperparameters $\omega_1$, $\omega_2$, and $\eta$ control the weights of the regularization terms introduced in the previous sections.
	While maintaining image-oriented processing, such design incorporates domain-specific knowledge from wireless communications,  thereby enhancing the  interpretability of the neural network.

	In order to balance knowledge integration across the model and boost training efficiency, we also develop a three-stage training strategy to train different components of DCAT.
	Inspired by transfer learning, we configure distinct channel conditions across 3 stages, as detailed in Table \ref{tab:3-stages}, corresponding to progressively increasing difficulty levels in the task.  
	Each stage builds upon the “basic knowledge” acquired in the preceding stage.
    The wireless channel models for the first two stages can be formulated as “$Y \sneq X$” and “$Y \sneq X+N$”, primarily training the SwinT backbone to develop essential image compression and noise immunity abilities.
    The loss function does not require regularization terms during these stages.
	While in the final stage, we train the full DCAT including DC-attn and DC-permu under real dynamic channels to achieve comprehensive system knowledge integration.
	
	\vspace{1.2em} 
	To conclude, we thoroughly account for various factors of physical-layer transmission in practical systems, preventing potential inadaptability with real-world communication scenarios.
	Through the proposed neural network module and training methodology, DCAT can demonstrate the following strengths:
	\begin{enumerate}
		\item Robustness. DCAT is capable of maintaining robust performances under conditions of strong channel dynamics and severe selective fading, particularly when confronted with channel aging effects.
		\item Efficiency. Our scheme is able to achieve favorable cost-performance trade-off, instead of blindly increasing model parameters and computational overhead for performance gains. Given that networked intelligent applications typically require low end-to-end latency, the practical utility of DCAT is more prominent.
		\item Model interpretability. With the guidance of rule-based knowledge from wireless communications, DCAT is an interpretable and generalizable solution rather than merely fitting specific data distributions, which remains a pervasive challenge in data-driven approaches.
	\end{enumerate}
	\begin{table}[!t]
		\centering
		\caption{Explanation of the three-stage training strategy.}
		\label{tab:3-stages}
		\renewcommand{\arraystretch}{1.2}
		\setlength{\tabcolsep}{6pt}
		\begin{tabular}{@{}c|cc@{}}
			\specialrule{1.0pt}{0pt}{0pt} 
			\textbf{Stage} & Channel condition & Trained model\\
			\hline
			\ding{172}  & Noiseless  & w/o DC-attn \& DC-permu\\
			\hline
			\ding{173}  & AWGN  & w/o DC-attn \& DC-permu\\
			\hline
			\ding{174}  & Dynamic  & full DCAT\\
			\specialrule{1.0pt}{0pt}{0pt} 
		\end{tabular}
	\end{table}
	
	\section{Experimental Setup}\label{sec4}
	\subsection{Dataset and Channel Condition Settings}\label{sec4-1}
	For the wireless image transmission task over complex dynamic channels, we employ $3$-channel RGB images ($m\sneq3$) for experimental validation. We conduct model development and performance testing of DCAT alongside other baseline schemes based on the UDIS-D dataset \cite{UDIS-D}. It is a comprehensive collection which comprises 20,880 training and 2,212 test samples captured across diverse scenes and viewpoints.
	Furthermore, in order to validate the scalability of the approaches, we employ another dataset, PLACES365 \cite{PLACES365}, which contains tens of millions of images and exhibits more complex data distributions. We uniformly sample a subset from it to match the same training/testing set sizes as UDIS-D. During our experiment, all images are resized into the shape of $128\times128$.
	
	Meanwhile, we adopt the widely used channel generator QuaDRiGa \cite{QuaDRiGa} to simulate the complex dynamic channel of concern compliant with 3GPP standards \cite{3GPP}. We likewise construct two cases: the first (\textbf{A}) serves for models development and testing, while the second (\textbf{B}) evaluates scalability.
	Without loss of generality, we set the Tx to remain stationary, and configure several motion trajectories for the Rx. For each case, every trajectory lasts $4$ seconds with a simulation time unit of $3.125 \times 10^{-2}$ ms, where the Rx moves at different velocities.
	The parameter settings and corresponding CSI distortion caused by channel aging under different velocities are summarized in Table \ref{tab:quadriga}.
	
	\begin{table}[!t]
		\centering
		\renewcommand{\arraystretch}{1.2} 
		\caption{NMSE of aging CSI under different Rx velocities.}
		\label{tab:quadriga}
		\begin{tabular}{@{}c|ccccc@{}}
			\specialrule{0.8pt}{0pt}{0pt} 
			$\bm{v}$ (m/s) & 2  & 6 & 10 &15&21\\
			\hline
			NMSE  & 0.034  & 0.092 & 0.157 &0.471&1.134\\
			\specialrule{0.8pt}{0pt}{0pt} 
		\end{tabular}
		\par\smallskip\noindent
		\text{Case \textbf{A}: UMa Scenario, $f_c$=2.4 GHz, $T_p$=4.0 ms}
		
		\vspace{0.15cm}
		
		\begin{tabular}{@{}c|ccccc@{}}
			\specialrule{0.8pt}{0pt}{0pt} 
			$\bm{v}$ (m/s) & 2  & 6 & 10 &15&21\\
			\hline
			NMSE  & 0.013  & 0.072 & 0.267 &0.644&1.331\\
			\specialrule{0.8pt}{0pt}{0pt} 
		\end{tabular}
		\par\smallskip\noindent
		\text{Case \textbf{B}: RMa Scenario, $f_c$=5.0 GHz, $T_p$=2.0 ms}
	\end{table}

	\begin{table*}[!b]  
		\centering
		\caption{The hyper-parameters for network training.}
		\label{tab:hyper-parameters}
		\renewcommand{\arraystretch}{1.2}
		\setlength{\tabcolsep}{6pt}
		\begin{tabular}{@{}cc@{}}
			\specialrule{1.0pt}{0pt}{0pt} 
			\textbf{Config} & \textbf{Value} \\
			\specialrule{1.0pt}{0pt}{0pt} 
			Batch size & 64 \\ \hline
			Epochs & 120 for \textit{Stage \ding{172}}, 120 for \textit{Stage \ding{173}}, 240 for \textit{Stage \ding{174}} \\ \hline
			Optimizer & Adam \\ \hline
			Learning rate schedule & MultiStepLR (gamma=$[1/3,0.3]$) \\ \hline
			Milestones & $[25,90]$ for \textit{Stage \ding{172}}, $[25,90]$ for \textit{Stage \ding{173}}, $[50,180]$ for \textit{Stage \ding{174}}   \\ \hline
			Base learning rate & $3\times10^{-4}$ \\ \hline
			Regularization term weight & $\omega_1=0.5$,  $\omega_2=0.5$, $\eta=5.0$ \\ \hline
			\specialrule{1.0pt}{0pt}{0pt} 
		\end{tabular}
	\end{table*}
	
	\subsection{Performance Metric}\label{sec4-2}
	We select two evaluation metrics: the peak signal-to-noise ratio (PSNR) and the learned perceptual image patch similarity (LPIPS), which measure the quality from pixel-level distortion and holistic visual perception perspectives, respectively.
	
	Higher PSNR values indicate better image reconstruction quality, as defined by the following equation:
	\begin{equation}
		\label{PSNR-define}
		\text{PSNR}(\bm{S},\bm{\hat{S}}) = 10 \log_{10} \frac{(\max \bm{S})^2}{\text{MSE}(\bm{S},\bm{\hat{S}})} \text{ (dB)}
	\end{equation}
	where $\max \bm{S}$ denotes the maximum possible value of the original image $\bm{S}$ (255 for the 8-bit color). The MSE (Mean Square Error) is defined as $\text{MSE}(\bm{S},\bm{\hat{S}}) \sneq \frac{1}{3 H_0 W_0}\|\bm{S} -\bm{\hat{S}}\|_2^2$.
	
	LPIPS computes the dissimilarity between original and reconstructed images in a high-dimensional feature space through a pretrained deep neural network, such as AlexNet \cite{Alex} in our experiments. It is able to exhibit a superior alignment with human perception compared with conventional pixel-wise metrics \cite{LPIPS_CVPR}. Its value ranges between 0 and 1, with a smaller value indicating better performance, as expressed by:
	\begin{equation}
		\label{LPIPS-define}
		\text{LPIPS}(\bm{S},\bm{\hat{S}}) = \sum_{l=1}^{L_A} \frac{1}{H_l W_l} \sum_{h,w} \| \bm{w}_l \odot (\phi(\bm{S})^l-\phi(\bm{\hat{S}})^l) \|_2^2
	\end{equation}
	where $\phi(\bm{S})^l,\phi(\bm{\hat{S}})^l \in \mathbb{R}^{H_l \times W_l \times C_l}$ denote the intermediate features
	extracted from the $l$-th layer of the given AlexNet composed of a total of $L_A$ layers.
	In addition, $H_l \times W_l \times C_l$  represents its size, $\bm{w}_l \in \mathbb{R}^{C_l}$ is the weight vector.
	
	During training, to jointly optimize both metrics, the image perceptual loss in Eq. (\ref{total_loss}) is formulated as:
	\begin{equation}   
		\label{loss-perceptual}
		L_{image}= \mu_1 \text{MSE}(\bm{S},\bm{\hat{S}}) + \mu_2 \text{LPIPS}(\bm{S},\bm{\hat{S}})
	\end{equation}
	where $\mu_1 \sneq \frac{1}{72}\times (5\cdot \text{SNR} + 42)$ and $\mu_2 \sneq  \frac{85}{24}\times (-5\cdot \text{SNR} + 108)$ can dynamically adjust the optimization trade-off between the two metrics according to SNR variations.
	Under deteriorating channel conditions (lower SNR), greater emphasis is placed on holistic perceptual quality, whereas higher SNR conditions prioritize pixel-wise fidelity.

	\subsection{Benchmarks}\label{sec4-3}
	To validate the performance of our proposed DCAT, several deep learning-based JSCC approaches for wireless image transmission are implemented as baselines.

	\begin{itemize}
		\item \textbf{SwinT}: SwinT is an advanced backbone for image compression and transmission compared with CNN, ResNet, and ViT. 
		In our experimental comparisons, it serves as the ablation baseline for all other evaluated methods.
		\item \textbf{ADJSCC} \cite{ADJSCC}: ADJSCC can employ a single model for wireless image transmission across varying SNRs by incorporating a plug-and-play module named Attention Feature (AF). For fair comparison, we upgrade its original CNN backbone to SwinT.
		\item \textbf{WITT} \cite{WiTT}: WITT is another JSCC-based image transmission scheme designed with dynamic SNR adaptation. It achieves SOTA performance under simplified channel models like AWGN and Rayleigh fading.
		\item \textbf{SwinT-L} \cite{SwinT}: SwinT-Large Version (SwinT-L) solely retains the SwinT backbone, differing from other solutions in its increased number of Transformer layers according to \cite{SwinT}. Comparative analysis with it can better demonstrate the cost efficiency advantages.
	\end{itemize}

	To ensure fairness, the above baseline methods are trained based on the strategy described in Sec. \ref{sec3-4}, while employing the loss function specified in Eq.(\ref{loss-perceptual}).
	
	\subsection{Network Configuration}\label{sec4-4}
	In terms of model settings, we configure $M \sneq 4$ stages with symmetric network deployment of the $\mathrm{enc}_\alpha$ and $\mathrm{dec}_\beta$.
	We set $ [ C_1,C_2,C_3,C_4] \sneq [128,192,256,320]$ for the dimension of each stage and $ C \in \{64,128,192,256\}$ for different CRs. 
	For the SwinT backbone, we set $ [N_1,N_2,N_3,N_4] \sneq [2,2,2,2]$  for all other models except $ [N_1,N_2,N_3,N_4] \sneq [2,2,6,2]$ specifically for SwinT-L.
	
	We consider a channel SNR range of $[-6,18]$ dB, requiring all schemes to employ a single model adaptable to varying SNRs. During training, the complex dynamic channel follows the configuration of Case \textbf{A} in Sec. \ref{sec4-1} with uniformly distributed SNRs. While the tested SNR values are set to $\{-6,-3,0,3,6,9,12,15,18\}$ dB. Table \ref{tab:hyper-parameters} shows the hyper-parameters of network training based on the transfer learning strategy mentioned in Sec. \ref{sec3}. All models are implemented with PyTorch, with 4 NVIDIA GeForce RTX4090 GPUs.
	
	\section{Performance Evaluation and Analysis}\label{sec5}
	\begin{figure*}[!t]
		\centering
		\subfloat[PSNR vs. SNR (\textit{Aging Scenario with CP})]{\includegraphics[width=0.31\linewidth]{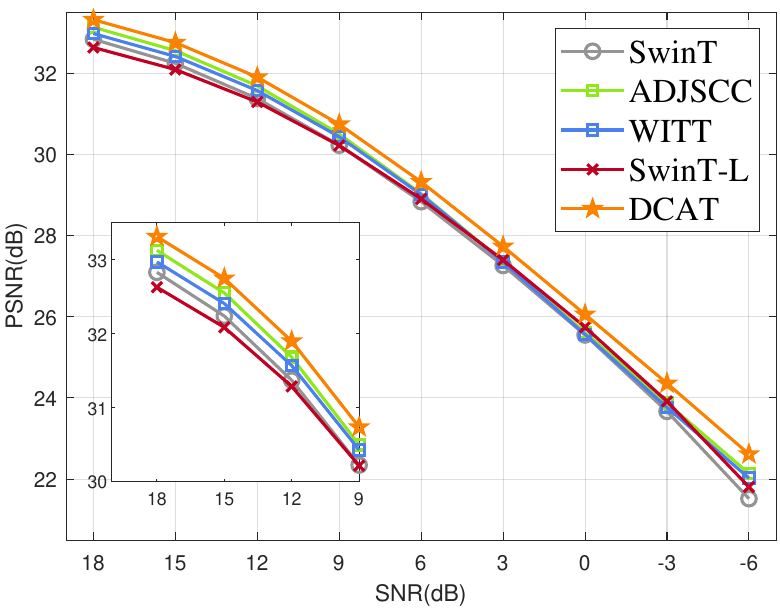}\label{fig:main-a}}
		\hfill
		\subfloat[PSNR vs. SNR (\textit{Aging Scenario})]{\includegraphics[width=0.31\linewidth]{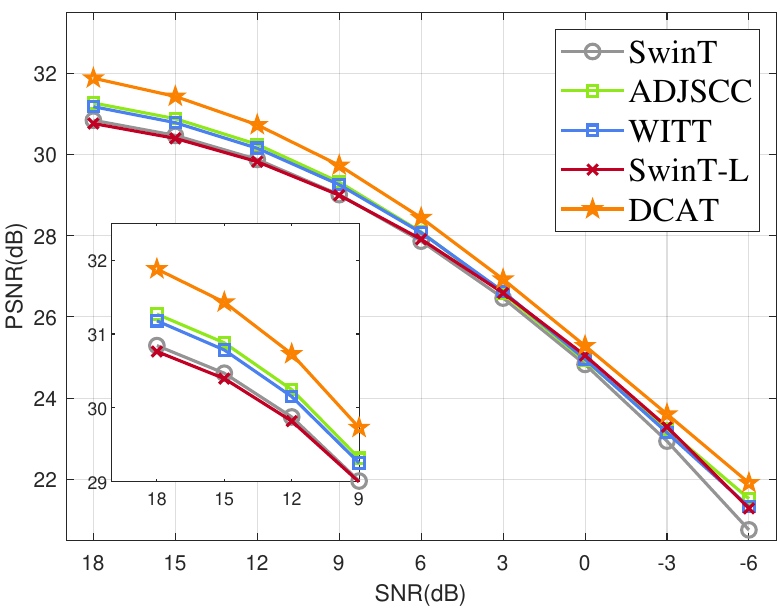}\label{fig:main-b}}
		\hfill
		\subfloat[PSNR vs. CR (\textit{Aging Scenario})]{\includegraphics[width=0.31\linewidth]{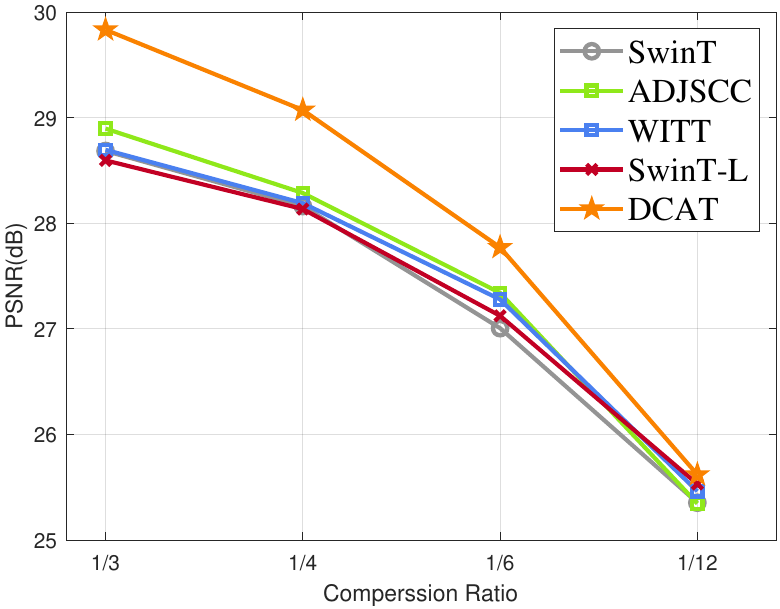}\label{fig:main-c}}
		\caption{Comparison of PSNR performance between the proposed DCAT and baseline schemes. (a) versus SNR under \textit{Aging Scenario with CP} (CR=$1/6$); (b) versus SNR under \textit{Aging Scenario} (CR=$1/6$); (c) versus CR under \textit{Aging Scenario} (averaged the metrics across all tested SNRs).}
		\label{pic-main-psnr}
	\end{figure*}
	\begin{figure*}[!t]
		\centering
		\subfloat[LPIPS vs. SNR (\textit{Aging Scenario with CP})]{\includegraphics[width=0.31\linewidth]{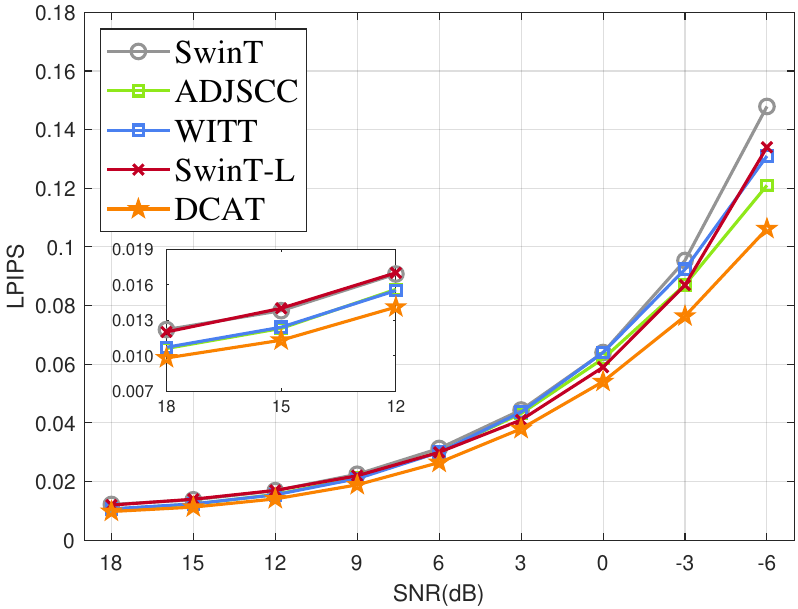}\label{fig:main-d}}
		\hfill
		\subfloat[LPIPS vs. SNR (\textit{Aging Scenario})]{\includegraphics[width=0.31\linewidth]{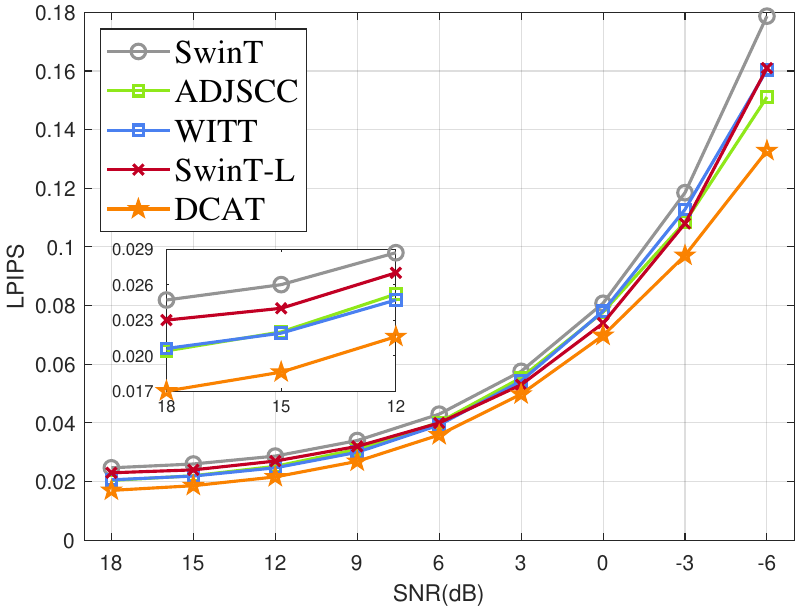}\label{fig:main-e}}
		\hfill
		\subfloat[LPIPS vs. CR (\textit{Aging Scenario})]{\includegraphics[width=0.31\linewidth]{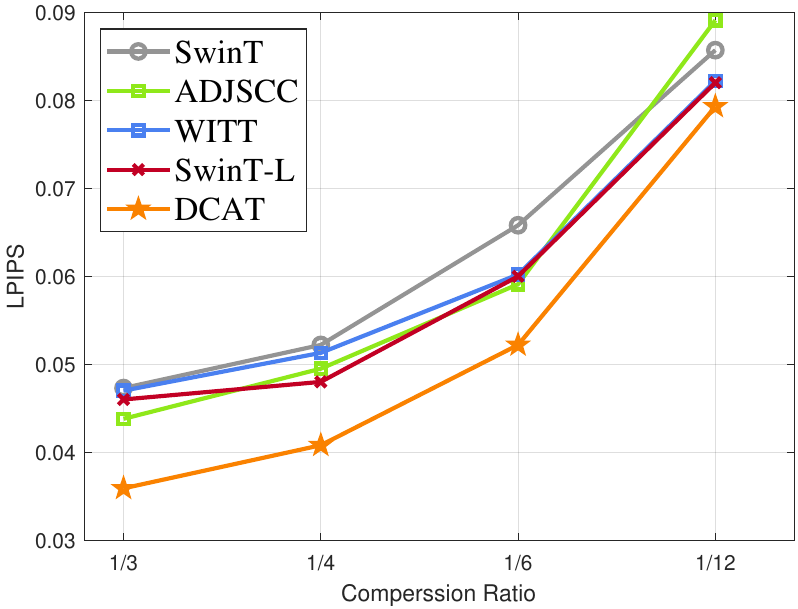}\label{fig:main-f}}
		\caption{Comparison of LPIPS performance between the proposed DCAT and baseline schemes. (a) versus SNR under \textit{Aging Scenario with CP} (CR=$1/6$); (b) versus SNR under \textit{Aging Scenario} (CR=$1/6$); (c) versus CR under \textit{Aging Scenario} (averaged the metrics across all tested SNRs).}
		\label{pic-main-lpips}
	\end{figure*}
	
	\subsection{Numerical Results Comparison with Baseline Schemes}\label{sec5-1}
	Fig. \ref{pic-main-psnr} compares the PSNR performance between our DCAT and baselines, with all schemes trained and tested on the UDIS-D dataset under Case \textbf{A} channel conditions as specified in Sec. \ref{sec4-1}.
	Fig. \ref{fig:main-a} and \ref{fig:main-b} present the PSNR versus channel SNR curves under \textit{Aging Scenario with CP} and \textit{Aging Scenario}, respectively. The results demonstrate the consistent superiority of our DCAT whether CP exists or not.
	The results reveal that channel aging degrades physical-layer symbol transmission quality, leading to overall performance degradation in Fig. \ref{fig:main-b} compared with \ref{fig:main-a}, while simultaneously amplifying the advantage of our proposed DCAT.
	Specifically, DCAT achieves the average PSNR gain of 0.324 dB (7.8\% improvement) over the suboptimal method under \textit{Aging Scenario with CP}, while the PSNR gain reaches 0.431 dB (10.5\% improvement) under \textit{Aging Scenario}.
	
	\begin{figure*}[!t]   
		\centering
		\setlength{\tabcolsep}{3pt}
		\renewcommand{\arraystretch}{0.5} 
		
		\newlength{\imgwidth}
		\setlength{\imgwidth}{\dimexpr (\linewidth - 3em - 21\tabcolsep)/6 \relax}
		
		\begin{tabular}{@{}m{0.5em}*{6}{c}@{}}
			& \textbf{Original image} & \textbf{SwinT} & \textbf{ADJSCC} & \textbf{WITT} & \textbf{SwinT-L} & \textbf{DCAT} \\[2pt] 
			
			\rotatebox[origin=c]{90}{\footnotesize\textbf{SNR=6 dB}} & 
			\raisebox{-.5\height}{\includegraphics[width=\imgwidth]{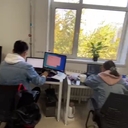}} & 
			\raisebox{-.5\height}{\includegraphics[width=\imgwidth]{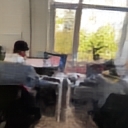}} & 
			\raisebox{-.5\height}{\includegraphics[width=\imgwidth]{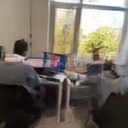}} & 
			\raisebox{-.5\height}{\includegraphics[width=\imgwidth]{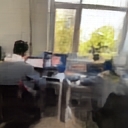}} & 
			\raisebox{-.5\height}{\includegraphics[width=\imgwidth]{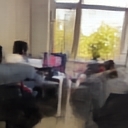}} & 
			\raisebox{-.5\height}{\includegraphics[width=\imgwidth]{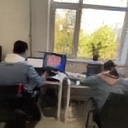}} \\[36pt]
			& \footnotesize \textbf{PSNR(dB) / LPIPS} & \footnotesize 24.00 / 0.1005 & \footnotesize 24.12 / 0.1204 & \footnotesize 24.24 / 0.1058 & \footnotesize 24.11 / 0.1235 & \footnotesize \textbf{28.38 / 0.0389} \\[3pt]
			
			\rotatebox[origin=c]{90}{\footnotesize\textbf{SNR=3 dB}} & 
			\raisebox{-.5\height}{\includegraphics[width=\imgwidth]{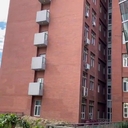}} & 
			\raisebox{-.5\height}{\includegraphics[width=\imgwidth]{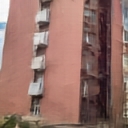}} & 
			\raisebox{-.5\height}{\includegraphics[width=\imgwidth]{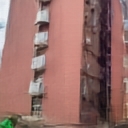}} & 
			\raisebox{-.5\height}{\includegraphics[width=\imgwidth]{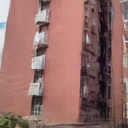}} & 
			\raisebox{-.5\height}{\includegraphics[width=\imgwidth]{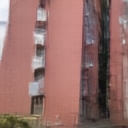}} & 
			\raisebox{-.5\height}{\includegraphics[width=\imgwidth]{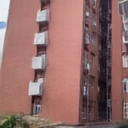}} \\[36pt]
			& \footnotesize \textbf{PSNR(dB) / LPIPS} & \footnotesize 24.49 / 0.1193 & \footnotesize 23.87 / 0.1134 & \footnotesize 24.32 / 0.0858 & \footnotesize 22.88 / 0.1535 & \footnotesize \textbf{26.60 / 0.0579} \\[3pt]
			
			\rotatebox[origin=c]{90}{\footnotesize\textbf{SNR=0 dB}} & 
			\raisebox{-.5\height}{\includegraphics[width=\imgwidth]{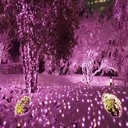}}& 
			\raisebox{-.5\height}{\includegraphics[width=\imgwidth]{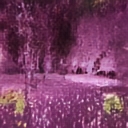}} & 
			\raisebox{-.5\height}{\includegraphics[width=\imgwidth]{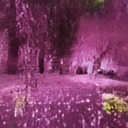}} & 
			\raisebox{-.5\height}{\includegraphics[width=\imgwidth]{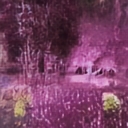}} & 
			\raisebox{-.5\height}{\includegraphics[width=\imgwidth]{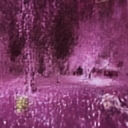}} & 
			\raisebox{-.5\height}{\includegraphics[width=\imgwidth]{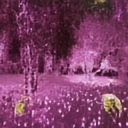}} \\[36pt]
			& \footnotesize \textbf{PSNR(dB) / LPIPS} & \footnotesize 19.31 / 0.2004 & \footnotesize 19.63 / 0.2068 & \footnotesize 19.61 / 0.2048 & \footnotesize 19.68 / 0.1934 & \footnotesize \textbf{21.68 / 0.0995} \\[3pt]
			
			\rotatebox[origin=c]{90}{\footnotesize\textbf{SNR=-3 dB}} & 
			\raisebox{-.5\height}{\includegraphics[width=\imgwidth]{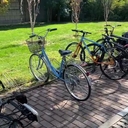}} & 
			\raisebox{-.5\height}{\includegraphics[width=\imgwidth]{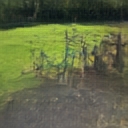}} & 
			\raisebox{-.5\height}{\includegraphics[width=\imgwidth]{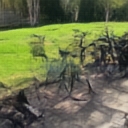}} & 
			\raisebox{-.5\height}{\includegraphics[width=\imgwidth]{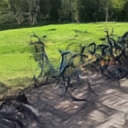}} & 
			\raisebox{-.5\height}{\includegraphics[width=\imgwidth]{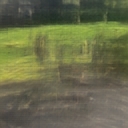}} & 
			\raisebox{-.5\height}{\includegraphics[width=\imgwidth]{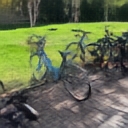}} \\[36pt]
			& \footnotesize \textbf{PSNR(dB) / LPIPS} & \footnotesize 15.61 / 0.3474 & \footnotesize 19.21 / 0.1636 & \footnotesize 19.28 / 0.1674 & \footnotesize 16.26 / 0.3962 & \footnotesize \textbf{20.71 / 0.1026} \\[3pt]
			
			\rotatebox[origin=c]{90}{\footnotesize\textbf{SNR=-6 dB}} & 
			\raisebox{-.5\height}{\includegraphics[width=\imgwidth]{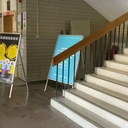}} & 
			\raisebox{-.5\height}{\includegraphics[width=\imgwidth]{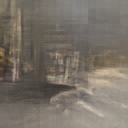}} & 
			\raisebox{-.5\height}{\includegraphics[width=\imgwidth]{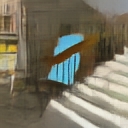}} & 
			\raisebox{-.5\height}{\includegraphics[width=\imgwidth]{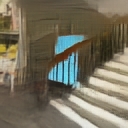}} & 
			\raisebox{-.5\height}{\includegraphics[width=\imgwidth]{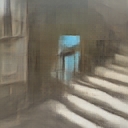}} & 
			\raisebox{-.5\height}{\includegraphics[width=\imgwidth]{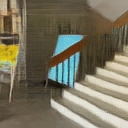}} \\[36pt]
			& \footnotesize \textbf{PSNR(dB) / LPIPS} & \footnotesize 14.75 / 0.4503 & \footnotesize 21.25 / 0.1305 & \footnotesize 20.87 / 0.1345 & \footnotesize 17.58 / 0.3211 & \footnotesize \textbf{22.95 / 0.0802} \\[3pt]

		\end{tabular}
		\caption{Visualization comparison under \textit{Aging Scenario}.
			All schemes under each SNR undergo the same channel aging condition with a fixed CR of $1/6$.}
		\label{pic_visualize}
	\end{figure*}

	Fig. \ref{fig:main-c} presents the CR-dependent performance variation under \textit{Aging Scenario}. The results clearly show DCAT maintains the best performance.
	As CR increases, the number of symbols per image transmission grows, leading to overall performance improvement while further amplifying the advantage of DCAT.
	When CR=$1/12$ ($C\sneq64$), the performance gain over the suboptimal scheme is 0.089 dB, while when CR=$1/3$ ($C\sneq256$), it reaches 0.936 dB.
	It occurs because higher CRs increase per-image transmission duration, thereby creating a larger optimization space to exploit time-selective fading and periodic channel aging characteristics in the system.
	
	Fig. \ref{pic-main-lpips} compares the LPIPS between DCAT and baselines. Similarly, Fig. \ref{fig:main-d} and \ref{fig:main-e} present the SNR-dependent performance variations under \textit{Aging Scenario with CP} and \textit{Aging Scenario}, while Fig. \ref{fig:main-f} demonstrates the CR-dependent LPIPS curve. 
	DCAT maintains the best performance across all test conditions, attaining LPIPS reduction of 0.0053 (10.9\%) and 0.0069 (12.6\%) under \textit{Aging Scenario with CP} and \textit{Aging Scenario} compared with the suboptimal baseline, respectively. The performance gain increases from 0.0029 (CR=$1/12$) to 0.0079 (CR=$1/3$) with longer image transmission duration.
	
	\subsection{Qualitative Results}\label{sec5-2}
	In this part, we analyze the qualitative results of wireless image transmission. Fig. \ref{pic_visualize} compares the reconstructed images of all baseline methods and DCAT with the original images. To ensure the visual effect, we select several compositionally complex and challenging samples from the UDIS-D dataset, while employing low SNR conditions for demonstration.
	
	We can easily observe that  baseline methods cannot show obvious performance improvements relative to their ablated counterpart SwinT, whereas DCAT consistently achieves lower distortion than them.
	DCAT excels at preserving fine-grained contours and textures (e.g., humans in the first row, bicycles in the forth row) and maintaining object structures even under severe channel conditions (SNR $<0$ dB), while baseline methods produce near-complete blurring.

	\subsection{Storage Overhead and Inference Complexity}\label{sec5-3}
	
	The model size, floating point operations (FLOPs), and inference time can indicate the model cost-effectiveness.  
	Thanks to its high-efficiency architecture, DCAT achieves the optimal performance without significant extra storage and computational cost.
	The overhead metrics are shown in Table \ref{tab:cost}.
	
	SwinT, serving as the ablated version for all schemes, yields the lowest metrics across all evaluations. Notably, WITT exhibits the highest parameter count due to its stacked FC layers in the SNR embedding module, while SwinT-L incurs the maximum computational FLOPs and inference time owing to its deeper network architecture. 
	Our DCAT does not achieve the highest value across all metrics, thereby its cost remains entirely within acceptable limits.
	Meanwhile, by explicitly incorporating system causality and physical-layer transmission information accessibility into our design, DCAT is fully capable of cost-effective deployment in practical applications.
	
	\begin{table}[!t]
		\centering
		\caption{Evaluation of complexity and computation cost per batch (batch size is set to 32) of each model.}
		\label{tab:cost}
		\renewcommand{\arraystretch}{1.35}
		\setlength{\tabcolsep}{5pt}
		\begin{tabular}{@{}c|ccccc@{}}
			\specialrule{1.0pt}{0pt}{0pt} 
			\textbf{Model} & \textbf{SwinT} & \textbf{ADJSCC} & \textbf{WITT} & \textbf{SwinT-L} & \textbf{DCAT} \\
			\specialrule{1.0pt}{0pt}{0pt} 
			Parameters (M) & 12.03 & 12.08 & 21.89 &  18.34 & 13.66 \\
			\hline
			FLOPs (G) & 210.12 & 210.13 & 217.25 &  261.90 & 210.18 \\
			\hline
			Inference time (ms) & 49.1 & 52.1 & 52.6 &  55.4 & 53.7 \\
			\specialrule{1.0pt}{0pt}{0pt} 
		\end{tabular}
	\end{table}

	\subsection{Ablation Study}\label{sec5-4}
	\begin{table*}[!t]
		\centering
		\caption{Ablation study across different velocities under \textit{Aging Scenario with CP}.}
		\label{tab:ab_ideal}
		\renewcommand{\arraystretch}{1.2}
		\setlength{\tabcolsep}{8pt}
		\begin{tabular}{|c|cc|cc|cc|cc|cc|}
			\specialrule{1.2pt}{0pt}{0pt} 
			\multicolumn{1}{|c|}{$v$ (m/s)} & \multicolumn{2}{c|}{2} & \multicolumn{2}{c|}{6} & \multicolumn{2}{c|}{10} & \multicolumn{2}{c|}{15} & \multicolumn{2}{c|}{21} \\ 
			\hline
			\multicolumn{1}{|c|}{Metric} & PSNR & LPIPS & PSNR & LPIPS & PSNR & LPIPS & PSNR & LPIPS & PSNR & LPIPS  \\ 
			\specialrule{1.0pt}{0pt}{0pt} 
			\textbf{DCAT} & \textbf{28.944} & \textbf{0.0367} & \textbf{28.809} & \textbf{0.0385} & \textbf{28.478} & \textbf{0.0431} & \textbf{28.493} & \textbf{0.0439} & \textbf{28.437} & \textbf{0.0444}  \\   
			\hline
			w/o DC-attn & 28.514 & 0.0438 & 28.360 & 0.0472 & 27.983 & 0.0539 & 27.997 & 0.0552 & 27.932 & 0.0558  \\  
			\hline
			w/o DC-permu & 28.940 & 0.0370 & 28.767 & 0.0393 & 28.411 & 0.0444 & 28.327 & 0.0462 & 28.124 & 0.0486  \\ 
			\specialrule{1.2pt}{0pt}{0pt} 
		\end{tabular}
	\end{table*}
	We perform ablation experiments to analyze the effect of different components within the DCAT model.  
	To elucidate the contributions of our design, ablation studies are conducted under both \textit{Aging Scenario with CP} and \textit{Aging Scenario}, with results presented in Table \ref{tab:ab_ideal} and \ref{tab:ab_aging}, respectively.  
	For each velocity level, performance metrics are averaged across all tested SNRs, while maintaining a fixed CR of $1/6$.
	
	\begin{table*}[!t]
		\centering
		\caption{Ablation study across different velocities under \textit{Aging Scenario}.}
		\label{tab:ab_aging}
		\renewcommand{\arraystretch}{1.2}
		\setlength{\tabcolsep}{8pt}
		\begin{tabular}{|c|cc|cc|cc|cc|cc|}
			\specialrule{1.2pt}{0pt}{0pt} 
			\multicolumn{1}{|c|}{$v$ (m/s)} & \multicolumn{2}{c|}{2} & \multicolumn{2}{c|}{6} & \multicolumn{2}{c|}{10} & \multicolumn{2}{c|}{15} & \multicolumn{2}{c|}{21} \\ 
			\hline
			\multicolumn{1}{|c|}{Metric} & PSNR & LPIPS & PSNR & LPIPS & PSNR & LPIPS & PSNR & LPIPS & PSNR & LPIPS  \\ 
			\specialrule{1.0pt}{0pt}{0pt} 
			\textbf{DCAT} & \textbf{28.431} & \textbf{0.0427} & \textbf{28.189} & \textbf{0.0464} & \textbf{27.749} & \textbf{0.0538} & \textbf{27.221} & \textbf{0.0620} & \textbf{26.351} & \textbf{0.0731}  \\ 
			\hline
			w/o DC-attn & 28.034 & 0.0484 & 27.790 & 0.0530 & 27.373 & 0.0607 & 26.971 & 0.0677 & 26.170 & 0.0789  \\
			\hline
			w/o DC-permu  & 28.303 & 0.0447 & 27.989 & 0.0491 & 27.477 & 0.0582 & 26.327 & 0.0784 & 25.448 & 0.0892  \\ 
			\hline
			w/o embed & 27.681 & 0.0533 & 27.239 & 0.0611 & 26.644 & 0.0724 & 25.838 & 0.0867 & 24.510 & 0.1126  \\ 
			\specialrule{1.2pt}{0pt}{0pt} 
		\end{tabular}
	\end{table*}
	Table \ref{tab:ab_ideal} presents the performance impact of removing either the DC-attn or DC-permute module compared with the full DCAT under \textit{Aging Scenario with CP}. It can be seen that DC-attn contributes significantly to performance gains, achieving an average PSNR improvement of 0.475 dB (11.6\%) and LPIPS reduction of 0.0099 (19.1\%). Whereas DC-permute shows relatively minor effects, gains of 0.118 dB (2.8\%) and 0.0018 (3.9\%) for PSNR and LPIPS, respectively.
	Furthermore, both modules exhibit positive correlation between the performance gain and the Rx velocity, particularly pronounced for DC-permute. It arises because the transceivers can acquire perfect CSI without distortion ($\bm{\hat{H}} \sneq \bm{H}$, $\bm{\tau_{ag}} \sneq 0$) with the help of CP.
	Correspondingly, the wireless channel influence is exclusively characterized by time-selective fading. Consequently, higher Rx velocities intensify channel selectivity, amplifying the quality disparity across different feature-channels, which significantly enhances the performance advantages brought by permutation module.
	
	The conclusions differ for \textit{Aging Scenario}, with detailed analyses provided according to Table \ref{tab:ab_aging}. 
	Firstly, DC-attn remains effective but demonstrates reduced efficacy compared with itself under \textit{Aging Scenario with CP}. Its performance gains, particularly in PSNR (0.397 dB at $v\sneq2$ m/s while 0.181 dB at $v\sneq21$ m/s), exhibit a negative correlation with the velocity. It occurs due to channel aging effects that degrade the reliability of CSI at transceivers, consequently impairing the global feature processing ability of attention mechanisms.
	However, DC-permute becomes much more effective as the distinct aging delay $\bm{\tau_{ag}}$ across feature-channels create greater inter-channel disparity, enabling the permutation mechanism to better preserve high-value features and reduce distortion.
	The last row “w/o embed” denotes the exclusion of physical-layer indication information embedding while preserving all submodules intact.
	This is implemented by forcing all relevant network inputs to zero values: $\overline{f_D}\sneq0$, $\bm{\tau_{ag}}\sneq0$.
	The results reveal that indication information (i.e., $\overline{f_D}$ and $\bm{\tau_{ag}}$,) proves critical for \textit{Aging Scenario}, which can improve PSNR by an average of 1.206 dB (32.5\%) and reduce LPIPS by 0.0216 (26.6\%).
	Additionly, the performance gain is more prominent when the velocity is high.
	Since increasing velocities exacerbate transmission impairments induced by dynamic wireless channels, this necessitates embedding indication information to guide the neural network to adapt, as mere consideration of SNR and CSI like other works proves insufficient.
	
	In summary, the results under different velocity levels further verify the effectiveness of our proposed scheme.
	Our reasonable network design significantly enhances the robustness against time-selective fading channels, particularly addressing channel aging effects.
	
	\subsection{DACT Scalability}\label{sec5-5}
	In Sec. \ref{sec3} we mentioned that due to the guidance of domain knowledge in wireless communications, DCAT exhibits excellent interpretability and is less prone to overfitting on specific data distributions. 
	In this part, we elaborate on this point from both source and channel perspectives. Specifically, all previous experiment results are obtained through training and testing on the UDIS-D image dataset under the QuaDRiGa channel configuration of Case \textbf{A}.
    As described in Sec. \ref{sec4-1}, we further validate the scalability of DCAT on more challenging data.
    \begin{figure}[!t] 
    	\centering
    	\subfloat[PSNR vs. Velocity]{
    		\includegraphics[width=0.46\linewidth]{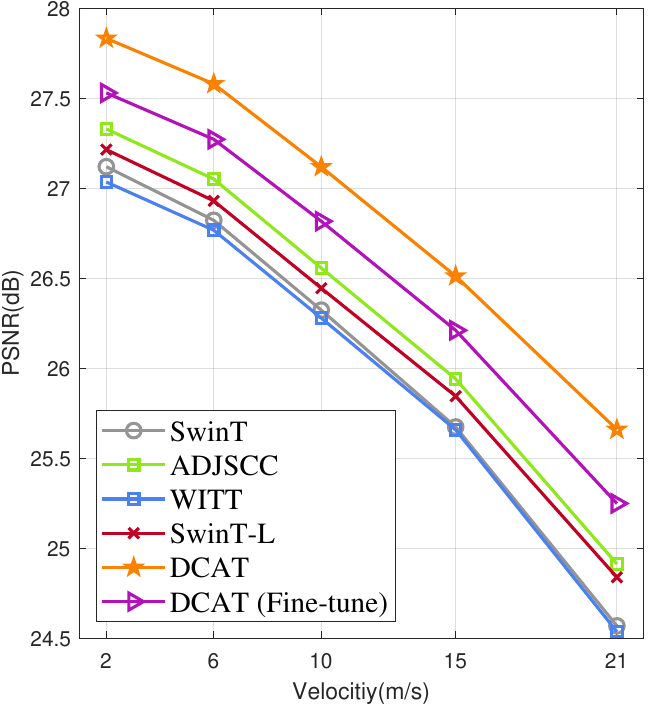} 
    		\label{pic_scalab_places:psnr}
    	}
    	\hfill 
    	\subfloat[LPIPS vs. Velocity]{
    		\includegraphics[width=0.46\linewidth]{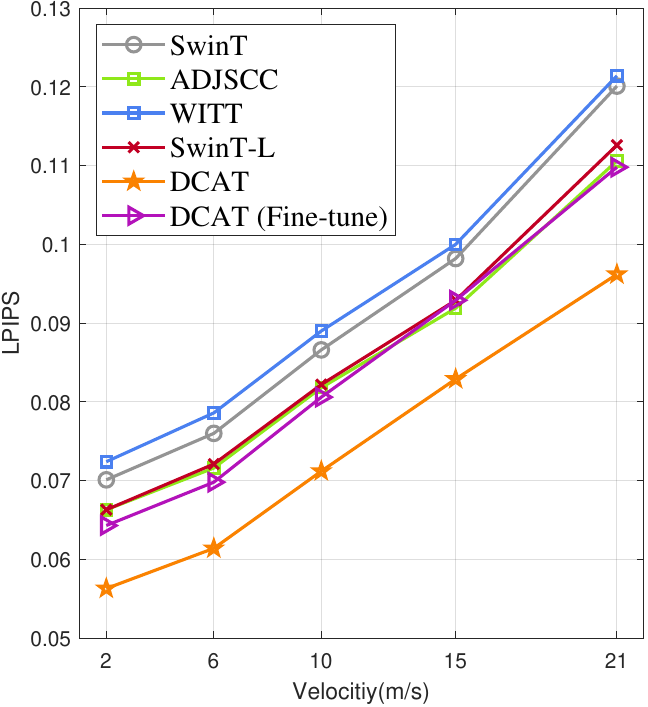} 
    		\label{pic_scalab_places:lpips}
    	}
    	\caption{The scalability experiment on the PLACES365 image dataset, averaged the metrics across all tested SNRs.}
    	\label{pic_scalab_places}
    \end{figure}

	When encountering unseen data, we first load the pre-trained parameters of all models, then retrain them for the same epochs. For DCAT framework, we adopt a fine-tuning scheme: the SwinT backbone remains frozen, while our proposed DC-attn, DC-permu and other lightweight modules including the convolutional layer, FCs, and normalization layers can be trained. Taking the model with the CR of $1/6$ as an example, only $3.06\text{M}/13.66\text{M}$ parameters are trainable, which can enable flexible adaptation to new scenarios with fully affordable fine-tuning overhead.
	
	Fig. \ref{pic_scalab_places} presents the validation results on PLACES365, with the x-axis representing the Rx velocity under \textit{Aging Scenario}. All other curves denote full-parameter training (full-shot), while the purple curve corresponds to the fine-tuned DCAT. Notably, DCAT full-shot achieves the best performance. 
	More importantly, the fine-tuned DCAT (average PSNR: 26.615 dB, average LPIPS: 0.0835) surpasses all baseline methods under full-shot learning (the optimal baseline ADJSCC yields 26.357 dB for PSNR and 0.0844 for LPIPS).
	
	\begin{figure}[!t] 
		\centering
		\subfloat[PSNR vs. Velocity]{
			\includegraphics[width=0.46\linewidth]{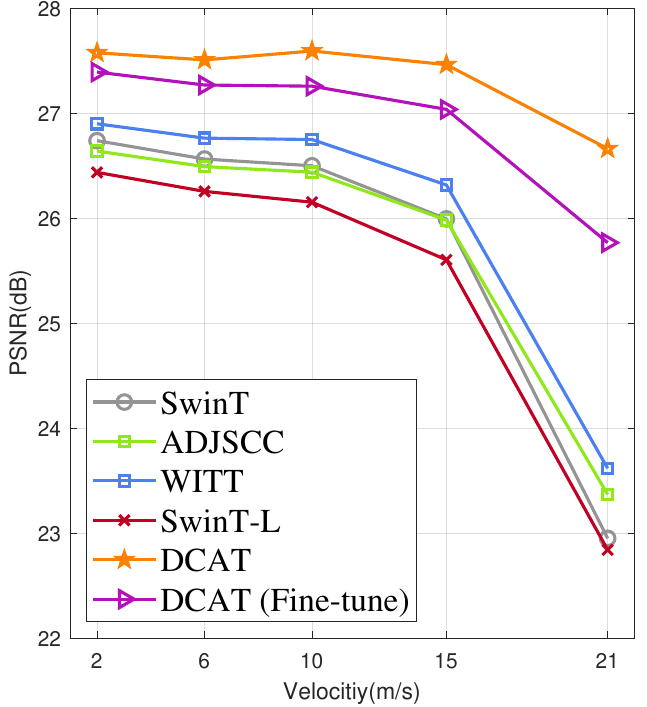} 
			\label{pic_scalab_quadriga:psnr}
		}
		\hfill 
		\subfloat[LPIPS vs. Velocity]{
			\includegraphics[width=0.46\linewidth]{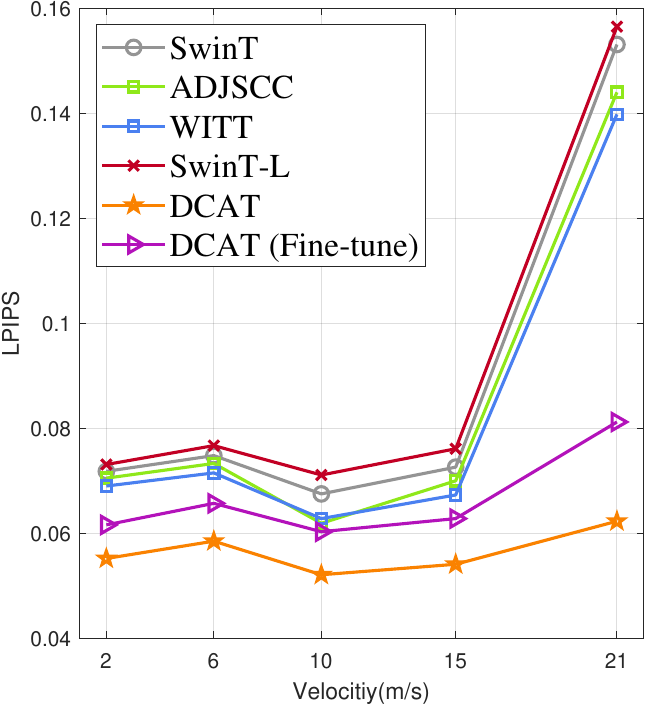} 
			\label{pic_scalab_quadriga:lpips}
		}
		\caption{The scalability experiment on the Case \textbf{B} channel configuration, averaged the metrics across all tested SNRs.}
		\label{pic_scalab_quadriga}
	\end{figure}
	
	Similarly, when encountering unseen channel conditions, DCAT demonstrates strong generalization capability, as illustrated in Fig. \ref{pic_scalab_quadriga}. Unlike Case \textbf{A} during model pre-training, Case \textbf{B} adopts a higher carrier frequency ($f_c$) and shorter pilot interval ($T_p$).
	Consequently, as the velocity exceeds a certain threshold, the combined effects of time-selective fading and channel aging significantly degrade transmission quality, leading to a rapid escalation in image distortion.
	It is obvious that DCAT effectively mitigates this issue, even its fine-tuned variant surpasses all baseline models retrained with full-shot learning.  At the velocity of $21$ m/s, the fine-tuned DCAT achieves a 2.149 dB PSNR improvement and 0.0586 LPIPS reduction compared with the top-performing baseline WITT, which is quite remarkable.
	
	The superior scalability and generalization of DCAT stem from its interpretable network design.  
	Unlike conventional approaches that treat neural networks as complete “black boxes”, our component design incorporates well-defined physical significance.
	Guided by domain-specific principles, DCAT achieves exceptional flexibility, empowering rapid deployment in practical applications with the minimal additional cost.
	
	\subsection{Transfer learning Efficacy}\label{sec5-6}
	This experiment aims to verify the efficacy of our training strategy in Sec. \ref{sec3-4}. aided DCAT for accommodating practical dynamic channels.
	We conducted two comparative groups to validate the benefit of transfer learning, as detailed in Table \ref{tab:transfer_learning_explain}. The “Dynamic” represents the actual operational scenario over the time-selective fading channel with aging.
	
	The “1-stage” approach involves direct training over the target dynamic channel , while the “2-stage” is first trained on noiseless channels. 
	The “3-stage” strategy is adopted by DCAT. 
	All strategies are ultimately evaluated on \textit{Aging Scenario}, with the total number of training epochs strictly controlled equal (480) to ensure equitable comparison.
	
\begin{figure}[!t] 
		\centering
		\subfloat[PSNR vs. SNR]{
			\includegraphics[width=0.46\linewidth]{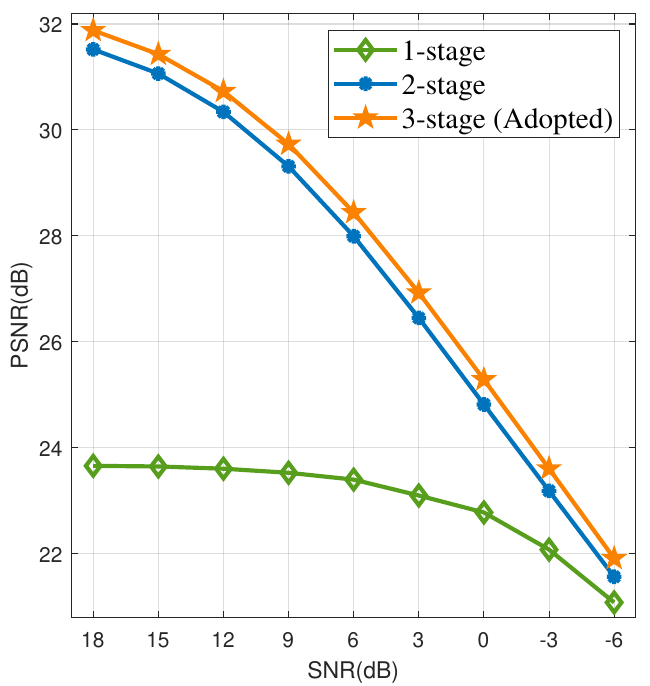} 
		}
		\hfill 
		\subfloat[LPIPS vs. SNR]{
			\includegraphics[width=0.46\linewidth]{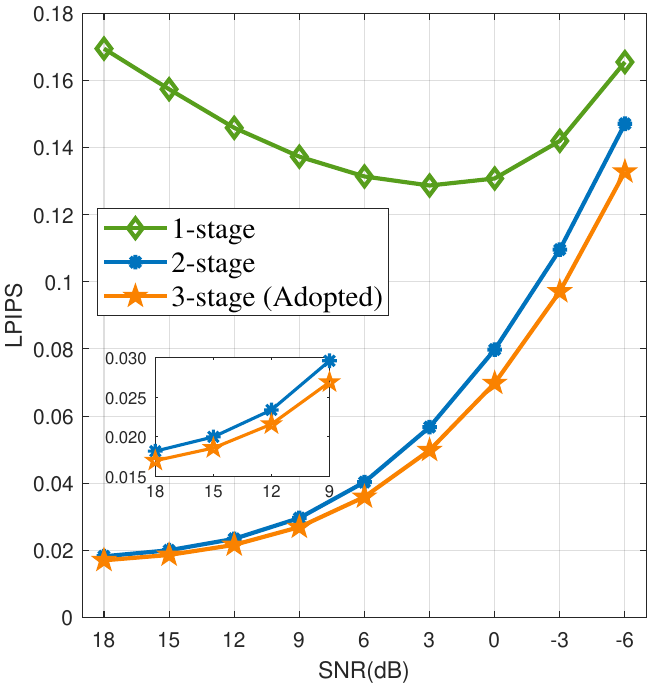} 
		}
		\caption{The performance of different transfer learning strategies. All tests are conducted under \textit{Aging Scenario} with CR =$1/6$.}
		\label{pic_transfer_learning}
	\end{figure}

	Both PSNR and LPIPS demonstrate the efficacy of the proposed transfer learning strategy as depicted in Fig. \ref{pic_transfer_learning}. The “1-stage” exhibits remarkably poor performance, even displaying anomalous trends in LPIPS. 
	This degradation stems from random parameter initialization combined with severe transmission impairments, which persistently hinder effective convergence of gradient-based optimization.
	
	The “2-stage” and “3-stage” ensure convergence of the SwinT backbone through initial training over the noiseless channel. Our adopted “3-stage” strategy further enhances model adaptation to additive noise via transfer learning over the AWGN channel, yielding better performance compared with the “2-stage”, particularly at low SNR conditions.
	
	We can conclude that transfer learning effectively decomposes a comprehensive and difficult task into several progressive subtasks.      
	Such “step by step” principle optimally balances knowledge between image feature processing and dynamic channel adaptation, forming an organic framework capable of handling image transmission over complex dynamic channels.
	
	\begin{table}[!t]
		\centering
		\caption{Illustration of channel condition settings for different training strategies.}
		\label{tab:transfer_learning_explain}
		\renewcommand{\arraystretch}{1.2}
		\setlength{\tabcolsep}{6pt}
		\begin{tabular}{!{\vrule width 1pt}c!{\vrule width 1pt}c|c|c!{\vrule width 1pt}}
			\hline
			Training strategy& Epoch 1-120 & Epoch 121-240 & Epoch 241-480  \\ \hline
			1-stage & \multicolumn{3}{c!{\vrule width 1pt}}{Dynamic} \\  \hline
			2-stage & \multicolumn{2}{c|}{Noiseless} & Dynamic\\ \hline
			3-stage (Adopted) & Noiseless& AWGN& Dynamic\\ \hline
		\end{tabular}
	\end{table}
	
	\section{Conclusion}\label{sec6}
	Wireless image transmission constitutes a critical research challenge since existing solutions lack consideration of physical-layer transmission in practical systems, limiting their real-world applicability.
	This paper proposes DCAT, an innovative image transmission scheme over complex dynamic channels, which can simultaneously possess the advantages of robustness, efficiency, and interpretability.
	Building upon the SwinT backbone, we integrate DC-attn and DC-permu to enable dynamic channel adaptive image transmission through physical-layer side information embedding.   
	Besides, by incorporating regularization methods and transfer learning, we effectively infuse knowledge from wireless communications, ensuring native compatibility with real-world systems.
	Experimental results demonstrate that DCAT outperforms existing JSCC approaches regardless of scenario or channel condition, while maintaining considerable scalability, empowering efficient deployment across diverse application domains.

	\fontsize{8pt}{8.3pt}\selectfont 
	
	\bibliographystyle{IEEEtran}
	\bibliography{myrefs.bib}
	

\end{document}